\documentclass[fleqn,10pt]{wlscirep}
\usepackage[utf8]{inputenc}
\usepackage[T1]{fontenc}
\usepackage{kotex}
\usepackage{xcolor}
\usepackage{subfigure}
\usepackage{epsfig}
\usepackage{authblk}

\title{Deep Interactive Learning-based ovarian cancer segmentation of H\&E-stained whole slide images to study morphological patterns of \textit{BRCA} mutation}
\author[1]{David Joon Ho}
\author[1]{M. Herman Chui}
\author[1]{Chad M. Vanderbilt}
\author[2]{Jiwon Jung}
\author[3]{Mark E. Robson}
\author[2]{Chan-Sik Park}
\author[4,6,*]{Jin Roh}
\author[5,6,*]{Thomas J. Fuchs}

\affil[1]{Department of Pathology, Memorial Sloan Kettering Cancer Center, New York, NY, USA}
\affil[2]{Department of Pathology, University of Ulsan College of Medicine, Asan Medical Center, Seoul, Republic of Korea}
\affil[3]{Department of Medicine, Memorial Sloan Kettering Cancer Center, New York, NY, USA}
\affil[4]{Department of Pathology, Ajou University School of Medicine, Suwon, Republic of Korea}
\affil[5]{Hasso Plattner Institute for Digital Health, Icahn School of Medicine at Mount Sinai, New York, NY, USA}
\affil[6]{The last two authors contributed equally.}
\affil[*]{jin.roh327@gmail.com; Thomas.Fuchs.AI@mssm.edu}

\begin{abstract}
Deep learning has been widely used to analyze digitized hematoxylin and eosin (H\&E)-stained histopathology whole slide images.
Automated cancer segmentation using deep learning can be used to diagnose malignancy and to find novel morphological patterns to predict molecular subtypes.
To train pixel-wise cancer segmentation models, manual annotation from pathologists is generally a bottleneck due to its time-consuming nature.
In this paper, we propose Deep Interactive Learning with a pretrained segmentation model from a different cancer type to reduce manual annotation time.
Instead of annotating all pixels from cancer and non-cancer regions on giga-pixel whole slide images, an iterative process of annotating mislabeled regions from a segmentation model and training/finetuning the model with the additional annotation can reduce the time.
Especially, employing a pretrained segmentation model can further reduce the time than starting annotation from scratch.
We trained an accurate ovarian cancer segmentation model with a pretrained breast segmentation model by 3.5 hours of manual annotation which achieved intersection-over-union of 0.74, recall of 0.86, and precision of 0.84.
With automatically extracted high-grade serous ovarian cancer patches, we attempted to train another deep learning model to predict \textit{BRCA} mutation.
The segmentation model and code have been released at \href{https://github.com/MSKCC-Computational-Pathology/DMMN-ovary}{https://github.com/MSKCC-Computational-Pathology/DMMN-ovary}.
\end{abstract}

\begin{document}

\flushbottom
\maketitle

\thispagestyle{empty}

\section*{Introduction} 
Deep learning, a subfield of machine learning, has shown an outstanding advancement in image analysis \cite{lecun2015, he2016} by training models using large public datasets \cite{russakovsky2015,krizhevsky2009,everingham2010,lin2014}.
Deep learning models have been used to analyze and understand digitized histopathology whole slide images to support some tedious and error-prone tasks \cite{fuchs2011,litjens2017,srinidhi2021,laak2021}.
For example, a deep learning model was used to identify breast cancers to search micrometastases and reduce review time \cite{steiner2018}.
Similarly, another deep learning model was used as a screening tool for breast lumpectomy shaved margin assessment to save time for pathologists by excluding the majority of benign tissue samples \cite{dalfonso2021}.
In addition, deep learning models have been investigated to discover novel morphological patterns indicating molecular subtypes from histologic images \cite{bera2019}.
Correlating digitized pathologic images with molecular information has contributed to prognosis prediction and personalized medicine \cite{komura2018}.
Specifically, molecular features from lung cancer \cite{coudray2018}, colorectal cancer \cite{kather2019,bilal2021}, and breast cancer \cite{wang2021,ektefaie2021,farahmand2022} can be predicted by deep learning models from hematoxylin and eosin (H\&E)-stained images.

All computational methods listed above either to diagnose cancers or to find biomarkers from cancer morphologies require accurate cancer segmentation of whole slide images.
Unlike common cancers where public datasets with annotation are provided \cite{bejnordi2018,bandi2019,amgad2019}, training deep learning-based segmentation models for rare cancers would require a vast amount of manual annotation which is generally time-consuming.
To overcome this challenge, we recently proposed Deep Interactive Learning (DIaL) to efficiently annotate osteosarcoma whole slide images to train a pixel-wise segmentation model \cite{ho2020}.
During an initial annotation step, annotators partially annotate tissue regions from whole slide images.
By iteratively training/finetuning a segmentation model and adding challenging patterns from mislabeled regions to the training set to improve the model, an osteosarcoma model was able to be trained by 7 hours of manual annotation.

In this paper, we develop an ovarian cancer segmentation model by 3.5 hours of manual annotation using DIaL.
The main contribution of this work to train an ovarian cancer segmentation model is to start DIaL from a pretrained triple-negative breast cancer (TNBC) model \cite{ho2021} to reduce manual annotation time by avoiding the initial annotation step.
Ovarian cancer accounts for approximately 2\% of cancer cases in the United States but is the fifth leading cancer causing death among women and the leading cause of death by cancer of the female reproductive system \cite{siegel2022}. 
High-grade serous ovarian cancer (HGSOC) is the most common histologic subtype and accounts for 70-80\% of deaths from all ovarian cancer \cite{lisio2019}.
Identifying \textit{BRCA1} or \textit{BRCA2} mutation status from HGSOC is important because family members of patients with germline mutations are at increased risk for breast and ovarian cancer and can benefit from early prevention strategies.
In addition, it offers increased treatment options for making therapeutic decisions. 
Deleterious variants in the \textit{BRCA1} or \textit{BRCA2} genes are strong predictors of response to poly ADP-ribose polymerase (PARP) inhibitors such as olaparib, niraparib, and rucaparib \cite{kim2015, ray2019}. 
Hence, the analysis of \textit{BRCA1/2} mutational status is crucial for individualized strategies for the management of patients with HGSOC.
Identification of \textit{BRCA1/2} mutation is currently done by genetic tests but some patients may not be able to get the genetic tests due to its high cost and limited resources. 
Although these limitations have been overcome by many factors including reduced \textit{BRCA} testing costs and next-generation sequencing, only 20\% of eligible women have accessed genetic testing in the United States \cite{childers2018}.
A cheaper approach to test \textit{BRCA1/2} mutation from H\&E-stained slides is desired to examine wider range of ovarian cancer patients and to provide proper treatments to them.
There has been an attempt to manually find morphological patterns of \textit{BRCA1/2} mutation \cite{soslow2012}.
In this study, we conduct deep learning-based experiments to screen \textit{BRCA1/2} mutation from cancer regions on H\&E-stained ovarian whole slide images automatically segmented by our model to potentially provide opportunities for more patients to examine their \textit{BRCA} mutational statuses.

\section*{Materials and methods}
\begin{figure*}[t!]
\centerline{\epsfig{figure=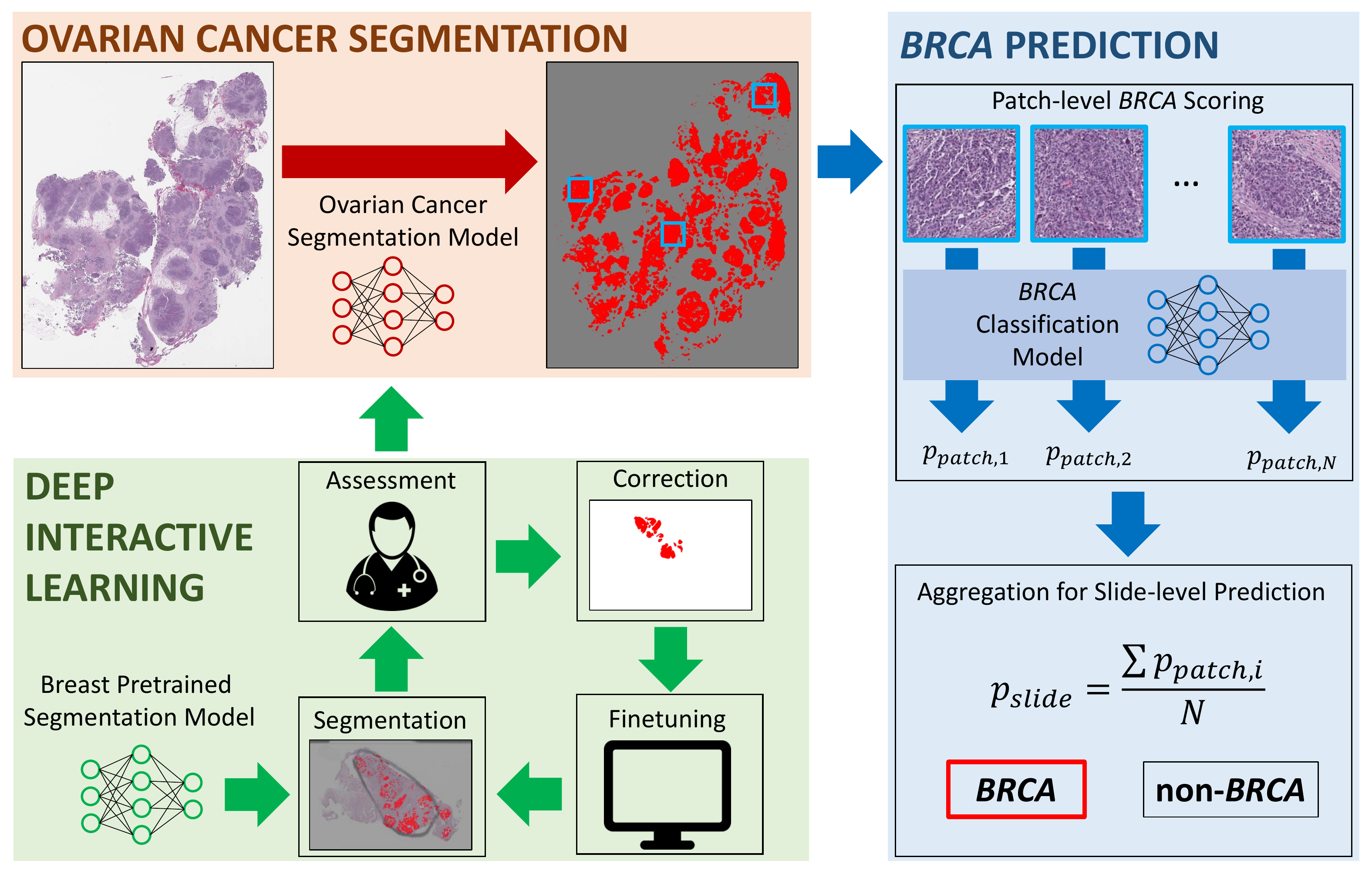,width=\textwidth}}
\caption{Block diagram of our proposed method to predict \textit{BRCA} mutation from H\&E-stained ovarian whole slide images. The first step was to segment ovarian cancer regions from whole slide images. To efficiently train the ovarian segmentation model, we used Deep Interactive Learning \cite{ho2020}. Since we started the process from a breast pretrained segmentation model, an annotator only spent 3.5 hours to annotate/correct whole slide images. After segmentation was done, cancer patches were processed by a \textit{BRCA} classification model to generate patch-level scores indicating the probability of \textit{BRCA} mutation. All patch-level scores were aggregated by averaging them to generate a slide-level prediction for \textit{BRCA} mutation.}
\label{fig:block}
\end{figure*}
Figure \ref{fig:block} shows the block diagram of our proposed method.
Our method is composed of two steps: (1) ovarian cancer segmentation and (2) \textit{BRCA} prediction.
The goal of this study is to predict \textit{BRCA} mutation from high-grade serous ovarian cancer, with an assumption that morphological patterns of the mutation would be shown on cancer regions.
Therefore, we trained a deep learning model for automated segmentation of ovarian cancer from H\&E-stained whole slide images to explore \textit{BRCA}-related patterns on segmented cancer regions.
Manual annotation process to train a deep learning-based segmentation model can be extremely time-consuming and tedious.
To reduce annotation time, we used Deep Interactive Learning we previously proposed \cite{ho2020}.
Our previous work had an initial annotation step to start manual annotation from scratch.
One contribution of this work is that we used a breast pretrained segmentation model \cite{ho2021} as our initial model to start Deep Interactive Learning to further reduce annotation time by avoiding initial annotation.
As a result, we were able to train our ovarian segmentation model with 3.5 hours of manual annotation.
After the segmentation model was trained, we attempted to predict \textit{BRCA} mutation based on cancer morphologies.
We trained another model using ResNet-18 \cite{he2016} to generate patch-level scores indicating the probabilities of \textit{BRCA} mutation.
The patch-level scores were aggregated by averaging all patch-level scores to generate a slide-level score to classify an input slide image to either \textit{BRCA} or non-\textit{BRCA}.
In this work, PyTorch \cite{paszke2019} was used for our implementation and an Nvidia Tesla V100 GPU was used for our experiments.

\subsection*{Data set}
To segment ovarian cancer and to predict \textit{BRCA} mutation based on tumor morphology, we collected 609 high-grade serous ovarian cancer cases at Memorial Sloan Kettering Cancer Center.
The MSK-IMPACT whole slide images were digitized in $20\times$ magnification by Aperio AT2 scanners.
Approximately 20\% of the cohort (119 images) have either \textit{BRCA1} or \textit{BRCA2} mutations and the other 80\% of the cohort (490 images) have no \textit{BRCA} mutation.
We randomly split 60\% of cases as a training set, 20\% as a validation set, and the remaining 20\% as a testing set, where the number of \textit{BRCA} and non-\textit{BRCA} images for training, validation, and testing are shown in Table \ref{tab:number_image}.
\begin{table}[ht]
\centering
{
\caption{The number of \textit{BRCA} and non-\textit{BRCA} whole slide images for our training, validation, and testing sets.}
\begin{tabular}{| c | c | c | c |}
	\hline
	 & \textit{BRCA} & Non-\textit{BRCA} & Total \\
	\hline
	Training Images & 73 & 294 & 367 \\
    \hline
    Validation Images & 23 & 98 & 121 \\
    \hline
    Testing Images & 23 & 98 & 121 \\
    \hline
    Total & 119 & 490 & 609 \\
    \hline
\end{tabular}
\label{tab:number_image}
}
\end{table}

\subsection*{Ovarian cancer segmentation}
\label{sec:segmentation}
We assume that morphological patterns caused by \textit{BRCA} mutation would most likely be present in cancer regions on ovarian whole slide images.
To train our \textit{BRCA} prediction model at scale, we trained an ovarian cancer segmentation model to automatically extract cancer regions and avoid any time-consuming manual segmentation.
In this work, we used Deep Multi-Magnification Network (DMMN) with multi-encoder, multi-decoder, and multi-concatenation \cite{ho2021} for ovarian cancer segmentation.
DMMN generates a segmentation patch in size of $256\times256$ pixels in $20\times$ based on various morphological features from a set of patches in size of $256\times256$ pixels from multiple magnifications in $20\times$, $10\times$, and $5\times$.

To train our DMMN model, manual annotation acquired from pathologists generally becomes a bottleneck.
Hence, we adopted Deep Interactive Learning (DIaL) \cite{ho2020} to reduce time for manual annotation.
As shown in Figure \ref{fig:block}, DIaL is composed of multiple iterations of segmentation, assessment, correction, and finetuning.
In each iteration, the annotators assess segmentation predictions generated by the previous model and correct any mislabeled regions.
The annotated patches are then included in a training set to train/finetune the segmentation model.
These iterations during DIaL help the annotators to efficiently annotate challenging morphological patterns so the training set can contain heterogeneous patterns of classes.
In our previous DIaL work, we started our initial annotation from scratch which took the majority of our annotation time.
To further reduce annotation time, we utilized a pretrained segmentation model from another cancer type to skip the initial annotation step.
In this work, we used a pretrained model to segment high-grade invasive ductal carcinoma from triple-negative breast cancer (TNBC) images \cite{ho2021} to train our model to segment high-grade serous ovarian carcinoma (HGSOC) because HGSOC and TNBC have shared morphological features such as large pleomorphic nuclei.

The pretrained model can segment 6 classes which are carcinoma, benign epithelium, stroma, necrosis, adipose tissue, and background.
During DIaL iterations, we kept the model to segment 6 classes but we converted benign epithelium, stroma, necrosis, adipose tissue, and background to be non-cancer to have binary segmentation.
The training set contained patches from both TNBC images and ovarian images.
To optimize the model, we used stochastic gradient descent (SGD) with a weighted cross entropy loss function, where a weight for class $c$, $w_c$, was determined by $w_c = 1-\frac{N_c}{N_t}$ where $N_c$ was the number of annotated pixels for class $c$ and $N_t$ is the total number of annotated pixels.
Random rotation, vertical and horizontal flips, and color jittering were used as data augmentation transformations \cite{buslaev2020}.
During the first training, we trained the model from randomly initialized parameters with a learning rate of $5\times10^{-5}$, a momentum of $0.99$, and a weight decay of $10^{-4}$, where the same hyperparameters were used from our previous breast segmentation training \cite{ho2021}.
After the first iteration, we finetuned the model from the previous parameters with a learning rate of $5\times10^{-6}$.
During training/finetuning iterations, we selected a model with the maximum intersection-over-union on a validation set as the final model of the iteration.
The final segmentation model processes patches on tissue regions extracted by Otsu Algorithm \cite{otsu1979}. 

\subsection*{\textit{BRCA} prediction}
With our assumption that \textit{BRCA} morphological patterns would be shown on cancer regions, we trained a patch-level classification model predicting \textit{BRCA} status from cancer patches.
Cancer patches were extracted from cancer masks from the ovarian cancer segmentation model.
A patch from a whole slide image was extracted if more than 50\% of pixels in the patch were segmented as cancer. 
It was observed that the number of cancer patches from training images was imbalanced.
If all cancer patches were used during training, morphological patterns on images with large cancer size would be more emphasized.
Therefore, we set an upper limit, $N^m$, as the maximum number of patches from a training whole slide image.
Specifically, if a whole slide image contained more than $N^m$ patches, $N^m$ training patches were randomly subsampled from the whole slide image.
Otherwise, all patches from the whole slide image were included in the training set.
In this work, we set $N^m_{BRCA} = 5000$ and $N^m_{nonBRCA}=1000$ where $N^m_{BRCA}$ was selected as the median value of the number of patches for \textit{BRCA} cases, and $N^m_{nonBRCA}$ was selected to balance the number of training patches between \textit{BRCA} and non-\textit{BRCA} classes.
Note that we did not subsample patches from the validation and testing sets to produce slide-level predictions based on all cancer regions on whole slide images.
The numbers of training, validation, and testing patches for \textit{BRCA} and non-\textit{BRCA} classes are shown in Table \ref{tab:number_patch}.
We trained three models in three different magnifications, $20\times$, $10\times$, and $5\times$.
We used ResNet-18 \cite{he2016} with patch size of $224\times224$ pixels.
We used the same weighted cross entropy as a loss function, Adam \cite{kingma2014} with learning rate of $10^{-5}$ as an optimizer, and used random horizontal and vertical flips, 90-degree rotations, and color jittering as data augmentation transformations \cite{buslaev2020}.

\begin{table}[ht]
\centering
{
\caption{The number of cancer patches for the training set, the validation set, and the testing set. Training patches between \textit{BRCA} and non-\textit{BRCA} are subsampled for balancing.}
\begin{tabular}{| c | c | c | c |}
	\hline
	 & \textit{BRCA} & Non-\textit{BRCA} & Total \\
	\hline
	Training Patches & 247,059 & 252,166 & 499,225 \\
    \hline
    Validation Patches & 120,753 & 592,232 & 712,985 \\
    \hline
    Testing Patches & 142,983 & 624,231 & 767,214 \\
    \hline
\end{tabular}
\label{tab:number_patch}
}
\end{table}

Although our prediction model generates patch-level scores, $p_{patch}$, the final goal of this work is to classify \textit{BRCA} mutation in slide-level.
Hence, we validated and tested our models in slide-level by aggregating patch-level scores.
In this work, a slide-level score, $p_{slide}$, was calculated by averaging the $N$ patch scores in an input whole slide image:
\begin{equation}
    p_{slide} = \frac{\sum p_{patch}}{N}
\end{equation}
We used area-under-curves (AUCs) as our evaluation metric.

\subsection*{Ethics declarations}
This study was approved by the Institutional Review Board at Memorial Sloan Kettering Cancer Center (Protocol \#21-473).

\section*{Results}
\subsection*{Training an ovarian cancer segmentation model with Deep Interactive Learning}
\begin{figure*}[ht!]
\centering
\subfigure[Image]{\frame{\epsfig{figure=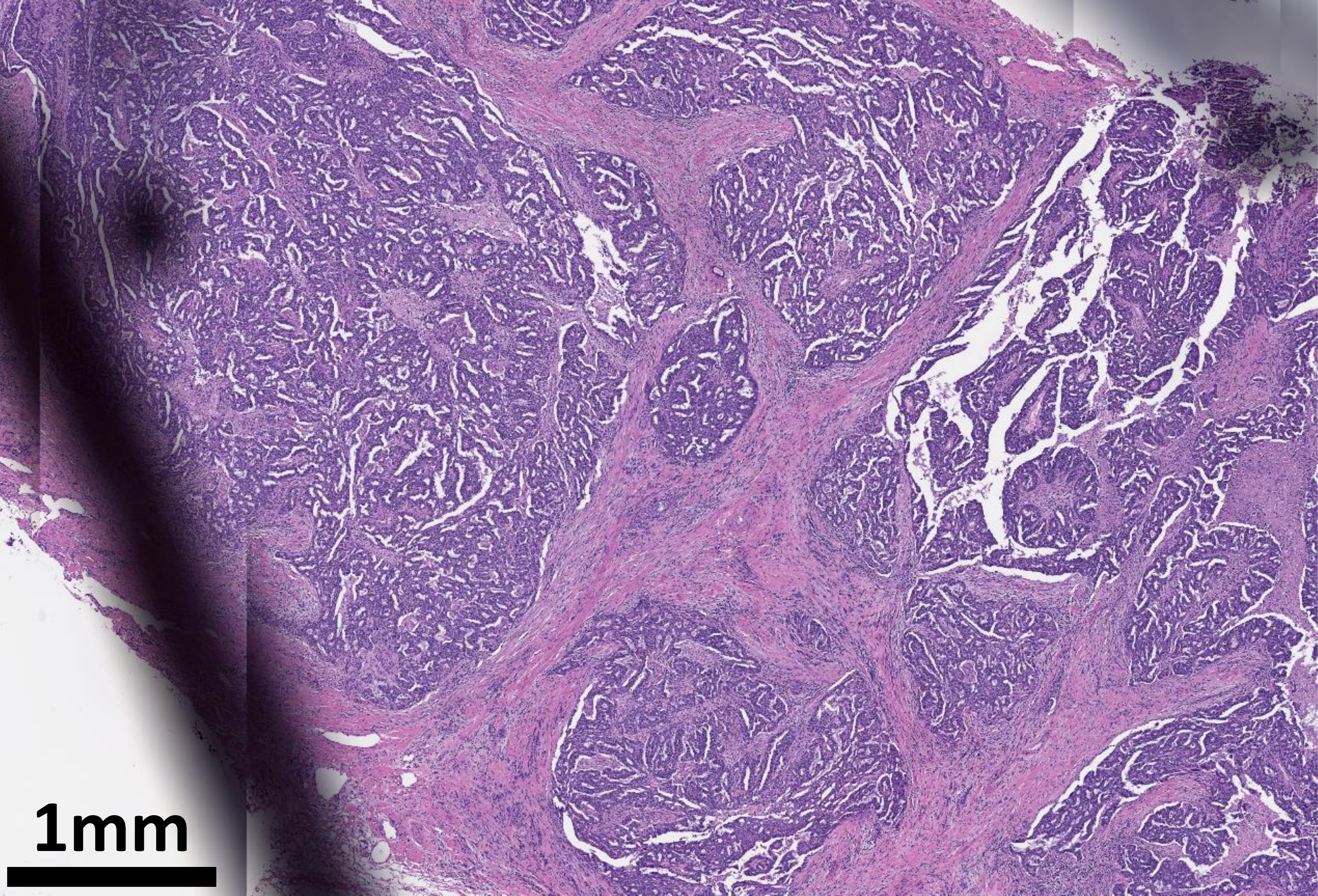,width = 0.32\textwidth}}}
\subfigure[Segmentation]{\frame{\epsfig{figure=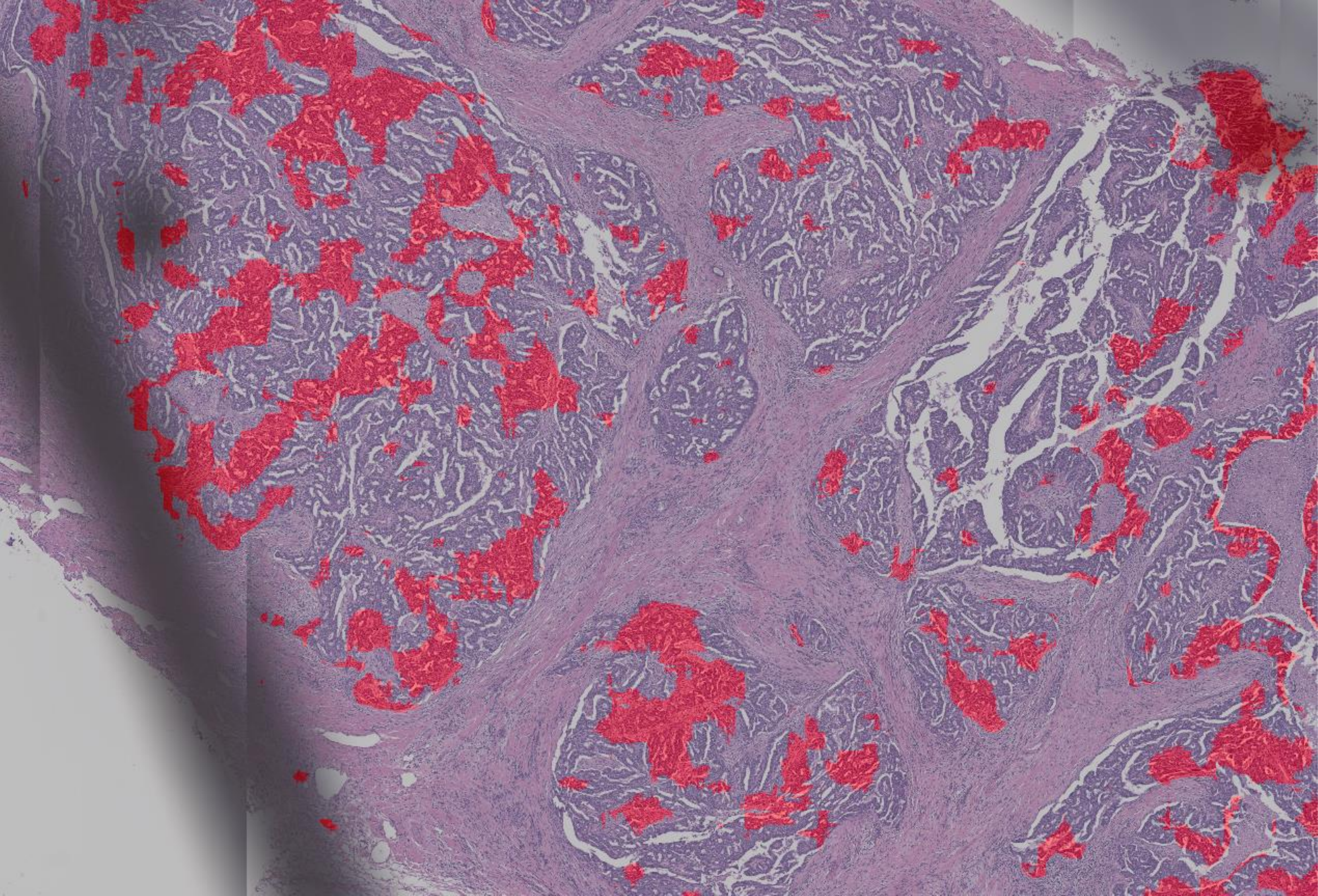,width = 0.32\textwidth}}}
\subfigure[Correction]{\frame{\epsfig{figure=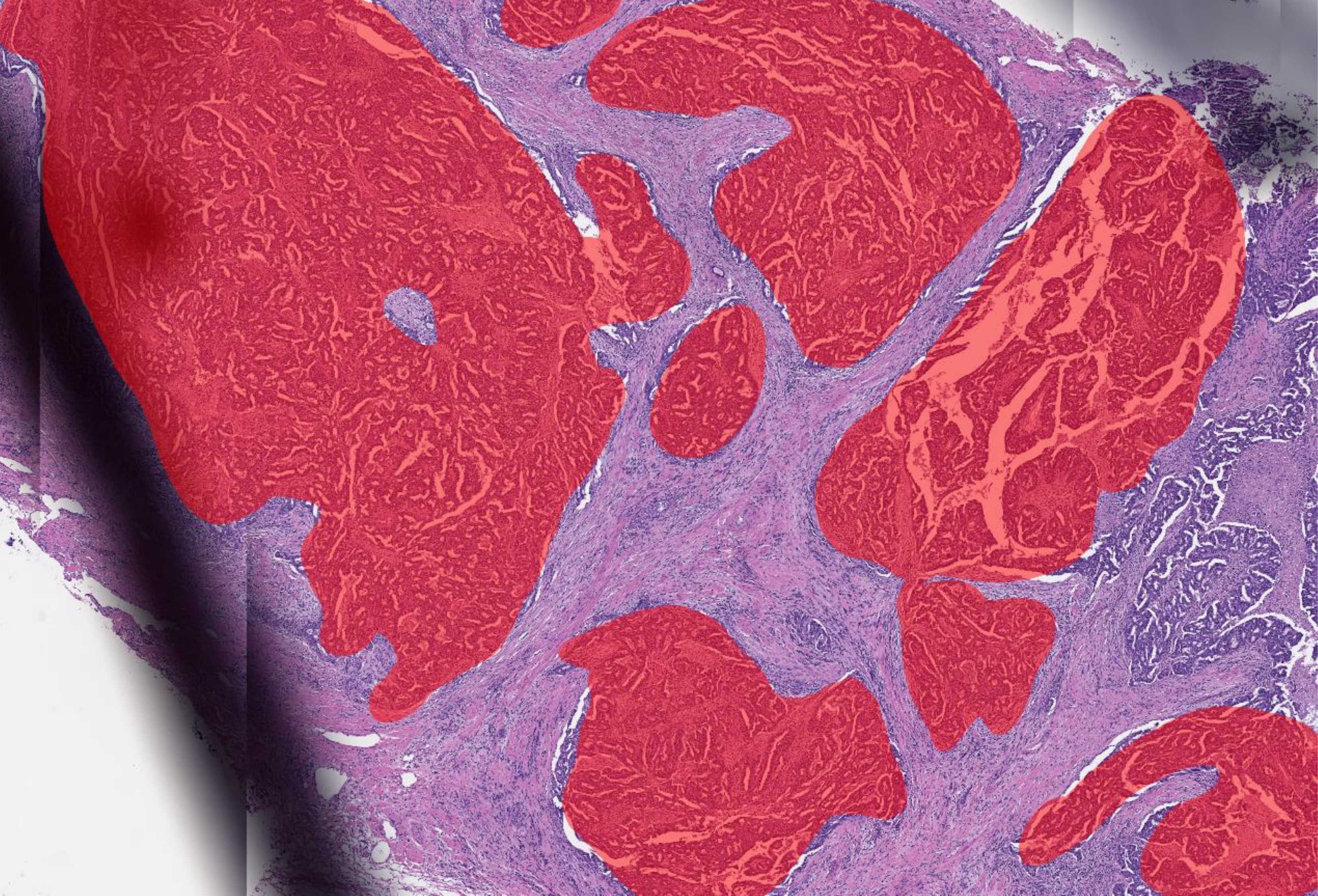,width = 0.32\textwidth}}}

\subfigure[Image]{\frame{\epsfig{figure=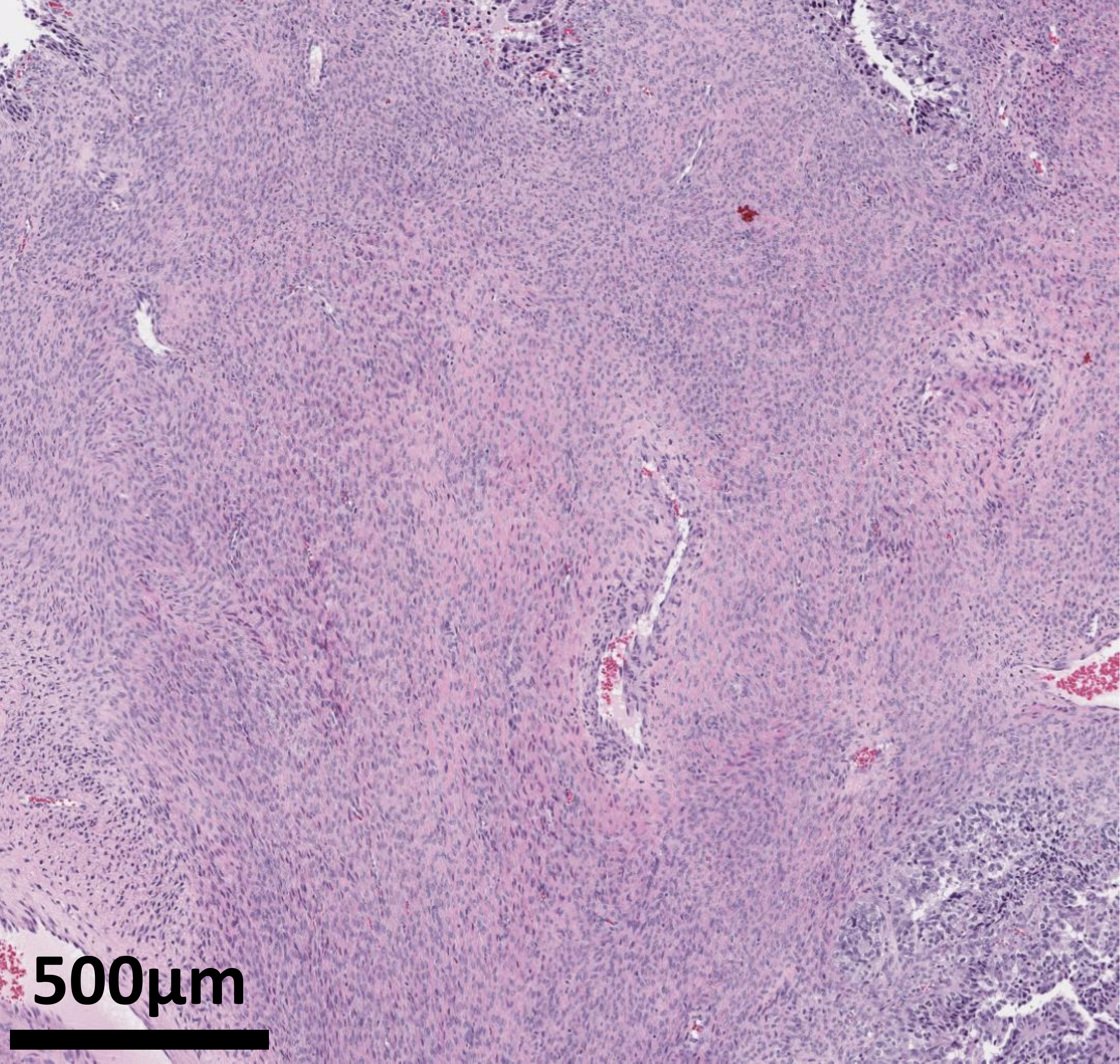,width = 0.32\textwidth}}}
\subfigure[Segmentation]{\frame{\epsfig{figure=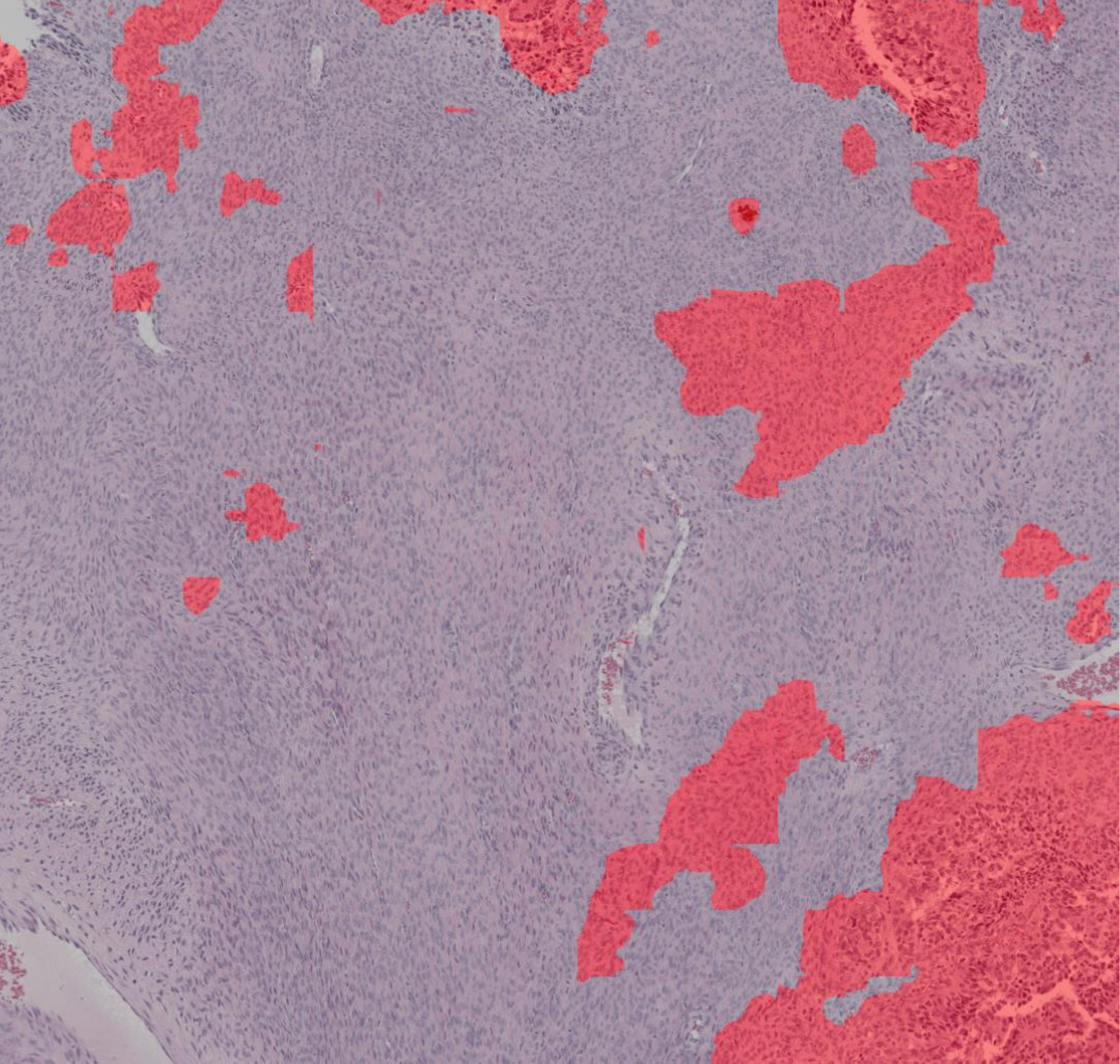,width = 0.32\textwidth}}}
\subfigure[Correction]{\frame{\epsfig{figure=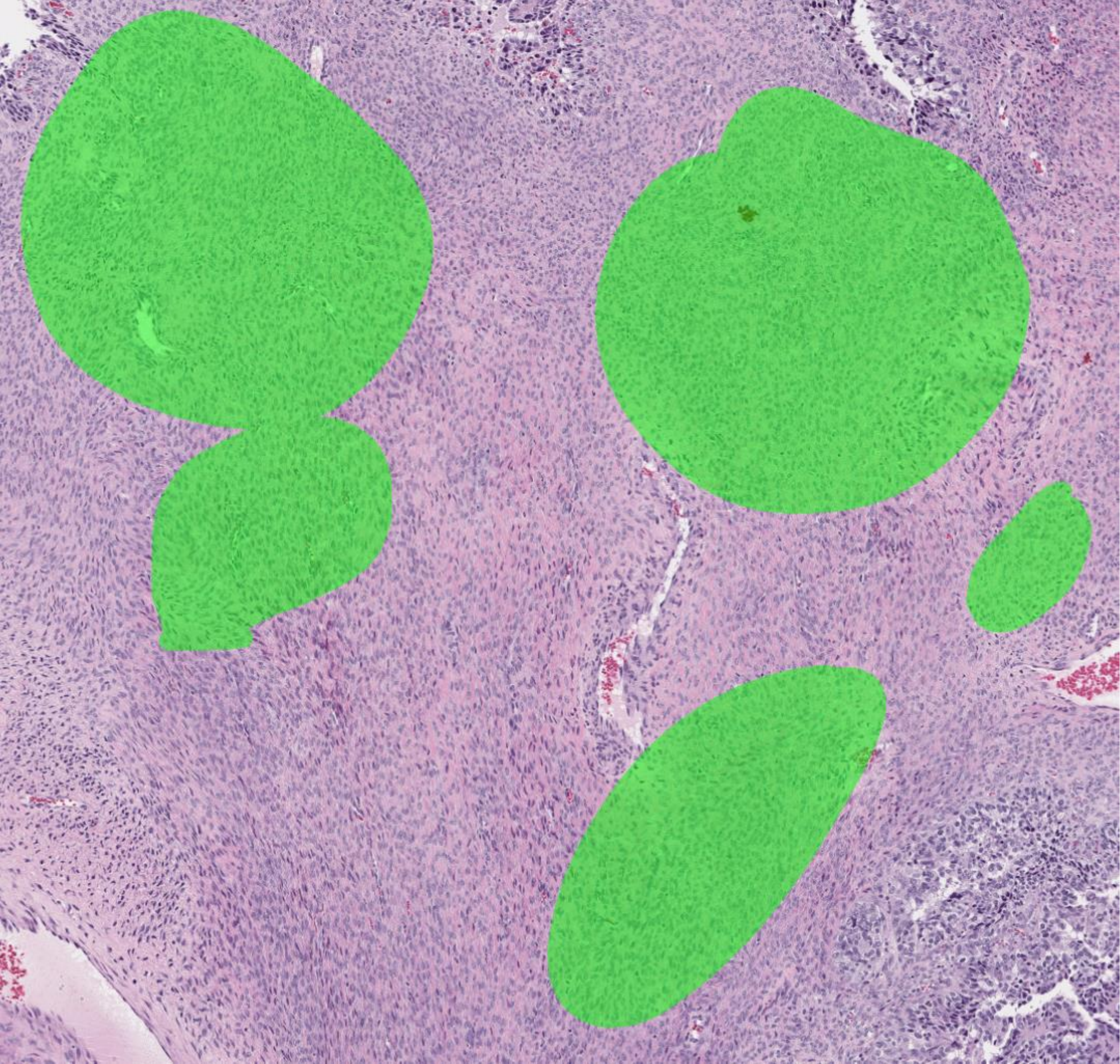,width = 0.32\textwidth}}}

\subfigure[Image]{\frame{\epsfig{figure=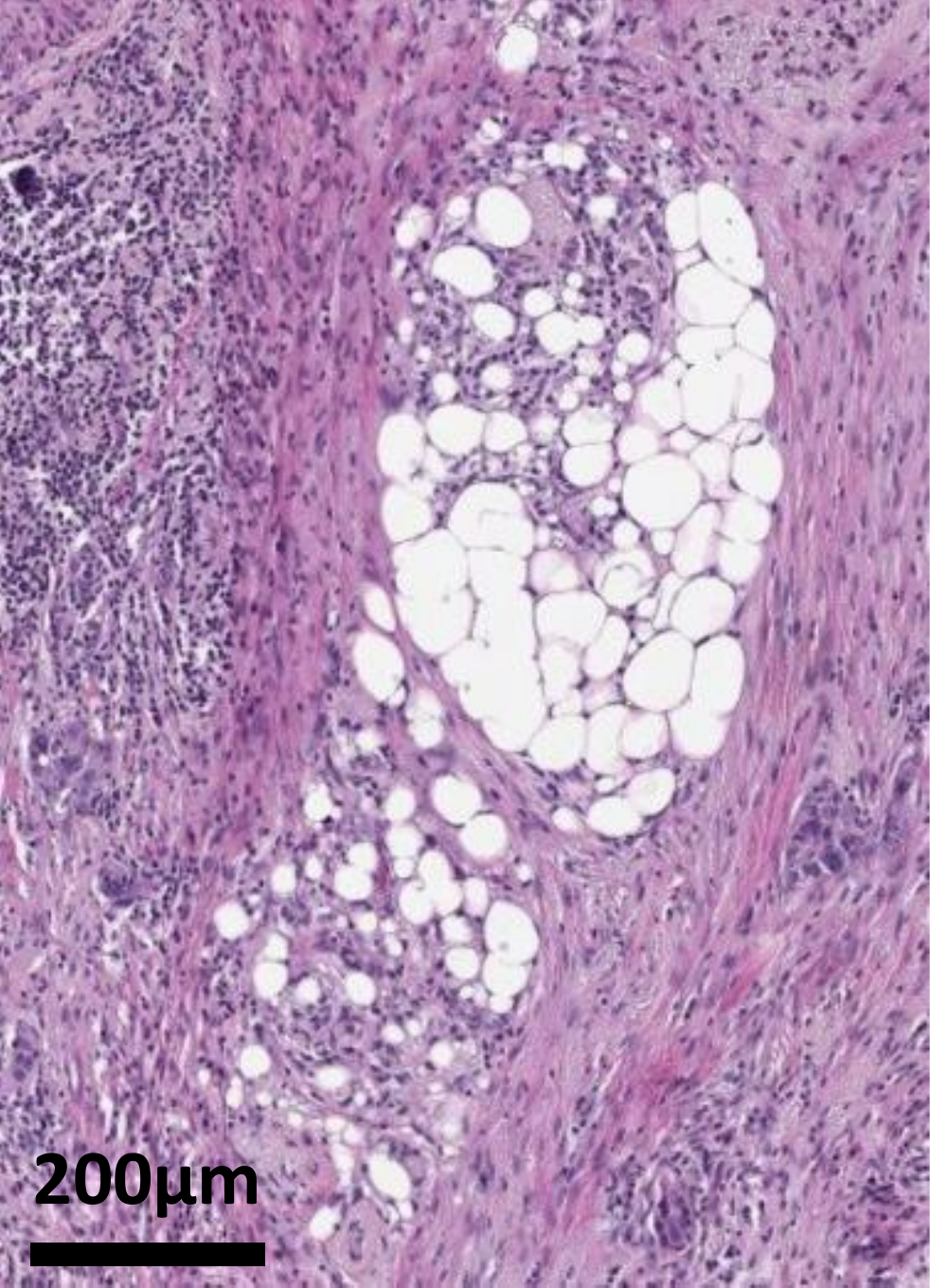,width = 0.32\textwidth}}}
\subfigure[Segmentation]{\frame{\epsfig{figure=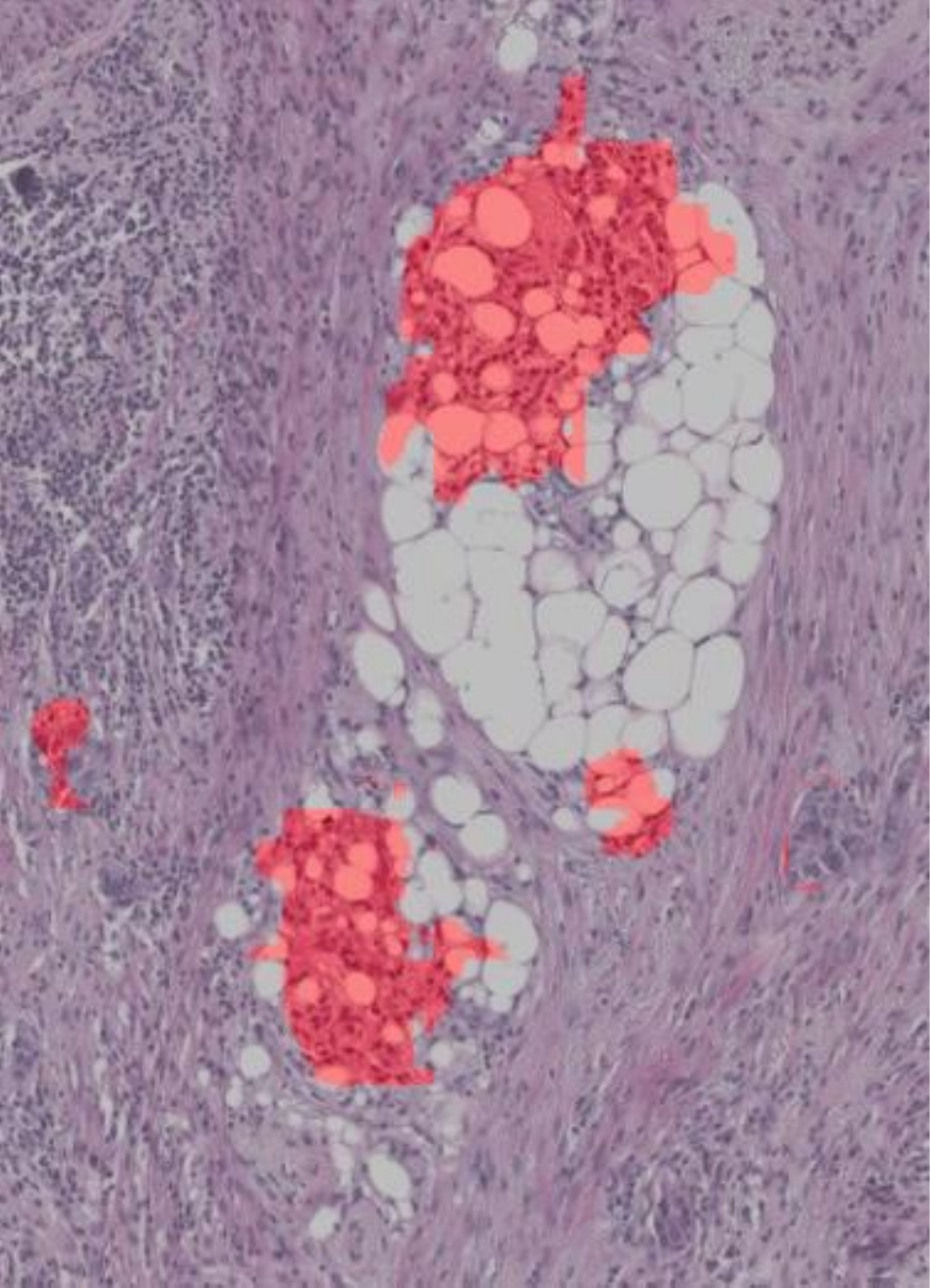,width = 0.32\textwidth}}}
\subfigure[Correction]{\frame{\epsfig{figure=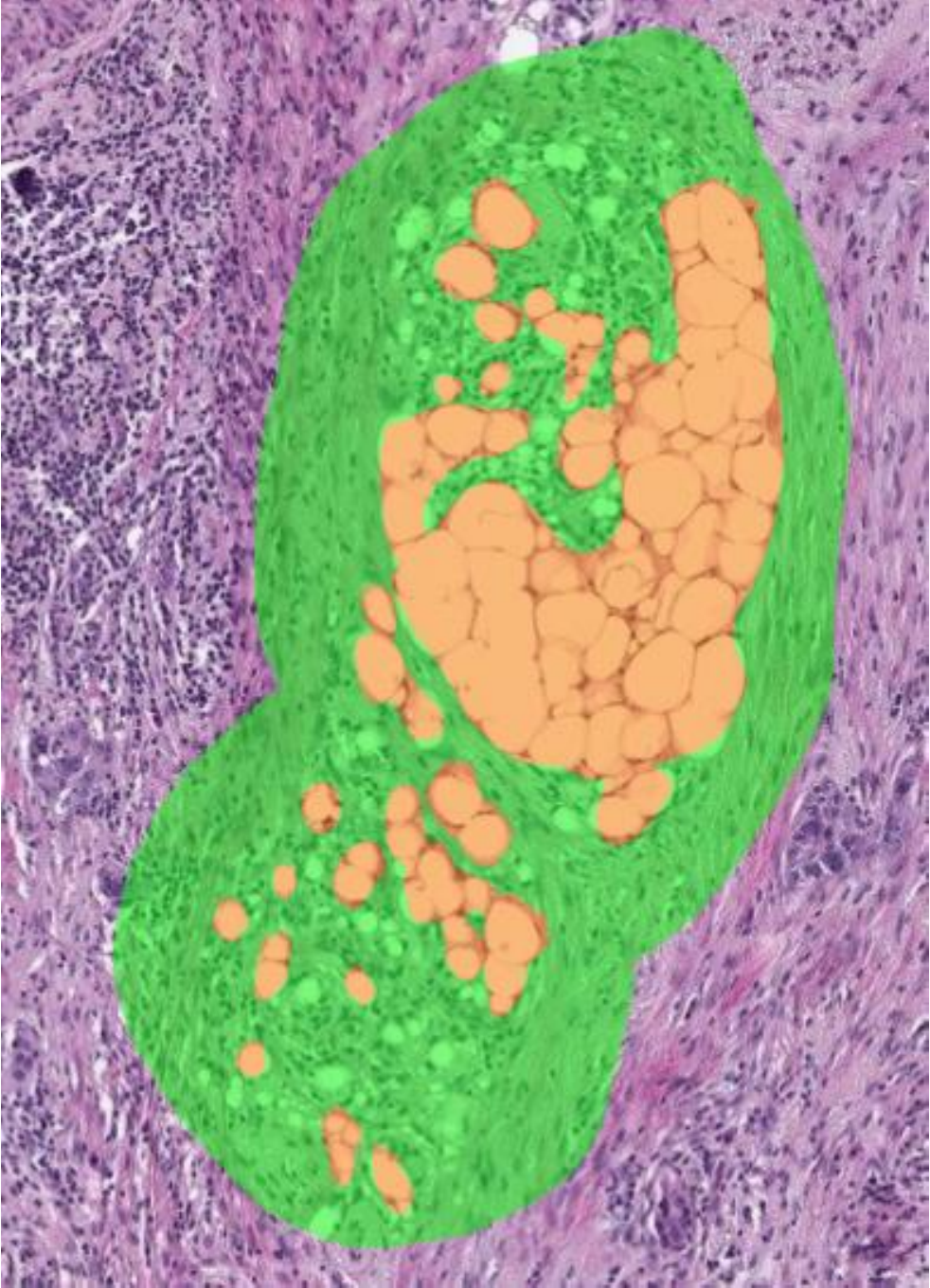,width = 0.32\textwidth}}}
\caption{Deep Interactive Learning for efficient annotation to train an ovarian cancer segmentation model. Instead of annotating all regions on ovarian whole slide images, the annotator can only annotate a subset of regions to train/finetune the ovarian cancer segmentation model. (a-c) The first iteration of correction of mislabeled cancer regions from $\mathcal{M}_0$. (d-f) The first iteration of correction of mislabeled stroma regions from $\mathcal{M}_0$. (g-i) The second iteration of correction of mislabeled fat necrosis regions from $\mathcal{M}_1$. Cancers are highlighted in red, stroma in green, and adipose tissue in orange.}
\label{fig:DIAL1}
\end{figure*}

We iteratively trained our ovarian cancer segmentation model using Deep Interactive Learning (DIaL) where annotation was done by an in-house slide viewer \cite{schuffler2021}.
We randomly selected 60 whole slide images from our training set where 20 images with \textit{BRCA1} status, 20 images with \textit{BRCA2} status, and 20 images with no \textit{BRCA} status were selected to make sure our ovarian cancer segmentation model can successfully segment all molecular subtypes.
After segmenting all 60 images using the pretrained breast model, denoted as $\mathcal{M}_0$, we observed papillary patterns for carcinoma and ovarian stroma were mislabeled because these patterns were not presented on triple-negative breast cancer images.
The annotator corrected mislabeled regions on 14 whole slide images to train the first model denoted as $\mathcal{M}_1$, which took approximately 1 hour.
During the second iteration, we observed that carcinoma and stroma were segmented correctly but some challenging patterns such as fat necrosis and fat cells in lymph node were mislabeled by $\mathcal{M}_1$.
The annotator looked for those challenging patterns in detail and annotated 11 whole slide images (3 images overlapping with the first correction step) to train the second model denoted as $\mathcal{M}_2$, took approximately 2 hours.
During the third iteration, we observed that some markers were mislabeled as carcinoma by $\mathcal{M}_2$, so the annotator corrected the other 3 whole slide images in 30 minutes to train the third model denoted as $\mathcal{M}_3$.
After finetuning the model, we observed $\mathcal{M}_3$ can successfully segment cancers on the training set so we completed the training stage.
In total, we annotated 25 ovarian whole slide images and spent 3.5 hours.
Figure \ref{fig:DIAL1} shows mislabeled regions by a segmentation model and corrected regions using DIaL.

\subsection*{Ovarian cancer segmentation evaluation}
\begin{figure*}[ht!]
\centering
\subfigure[Whole Slide Image]{\frame{\epsfig{figure=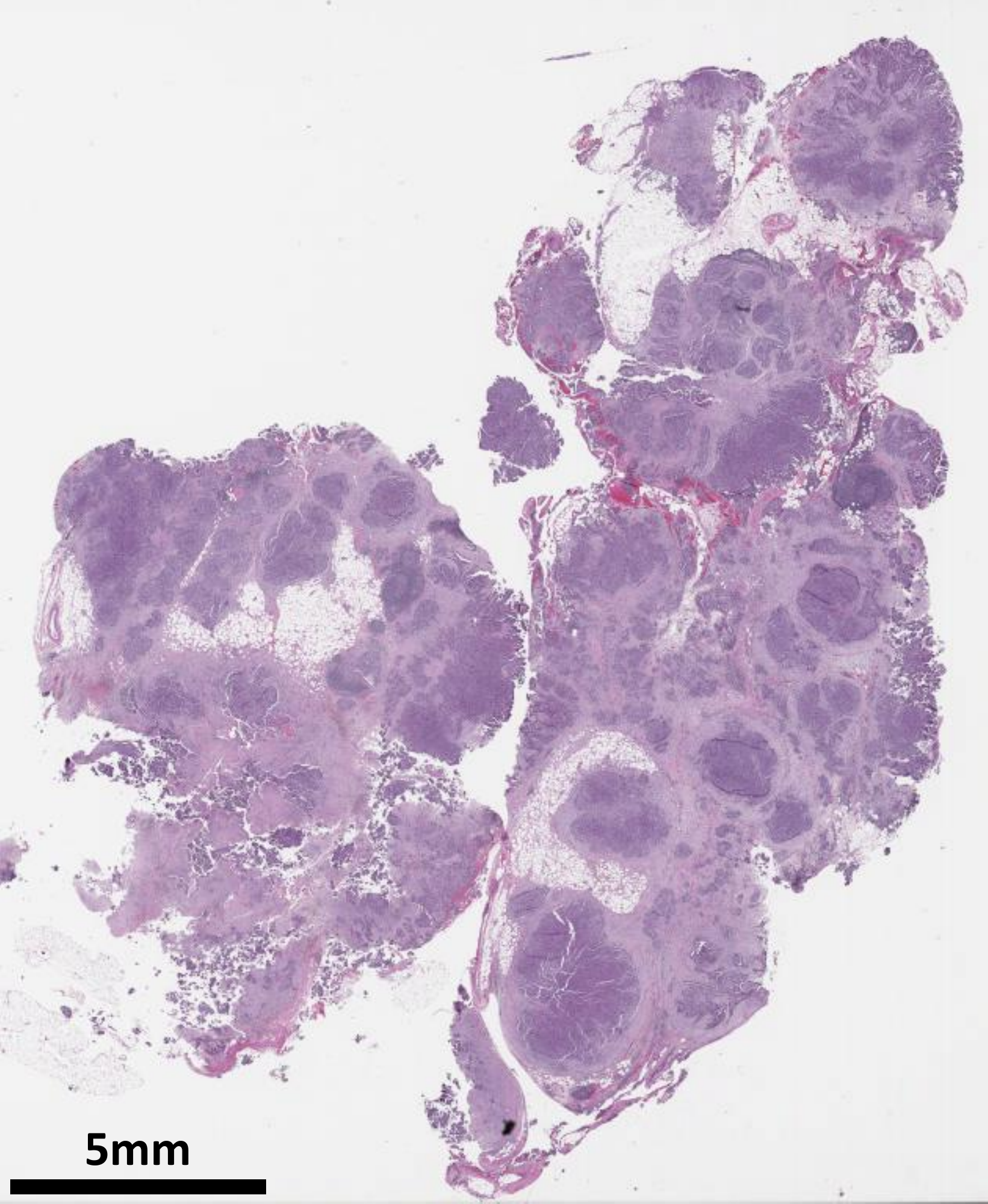,width = 0.32\textwidth}}}
\subfigure[Groundtruth]{\frame{\epsfig{figure=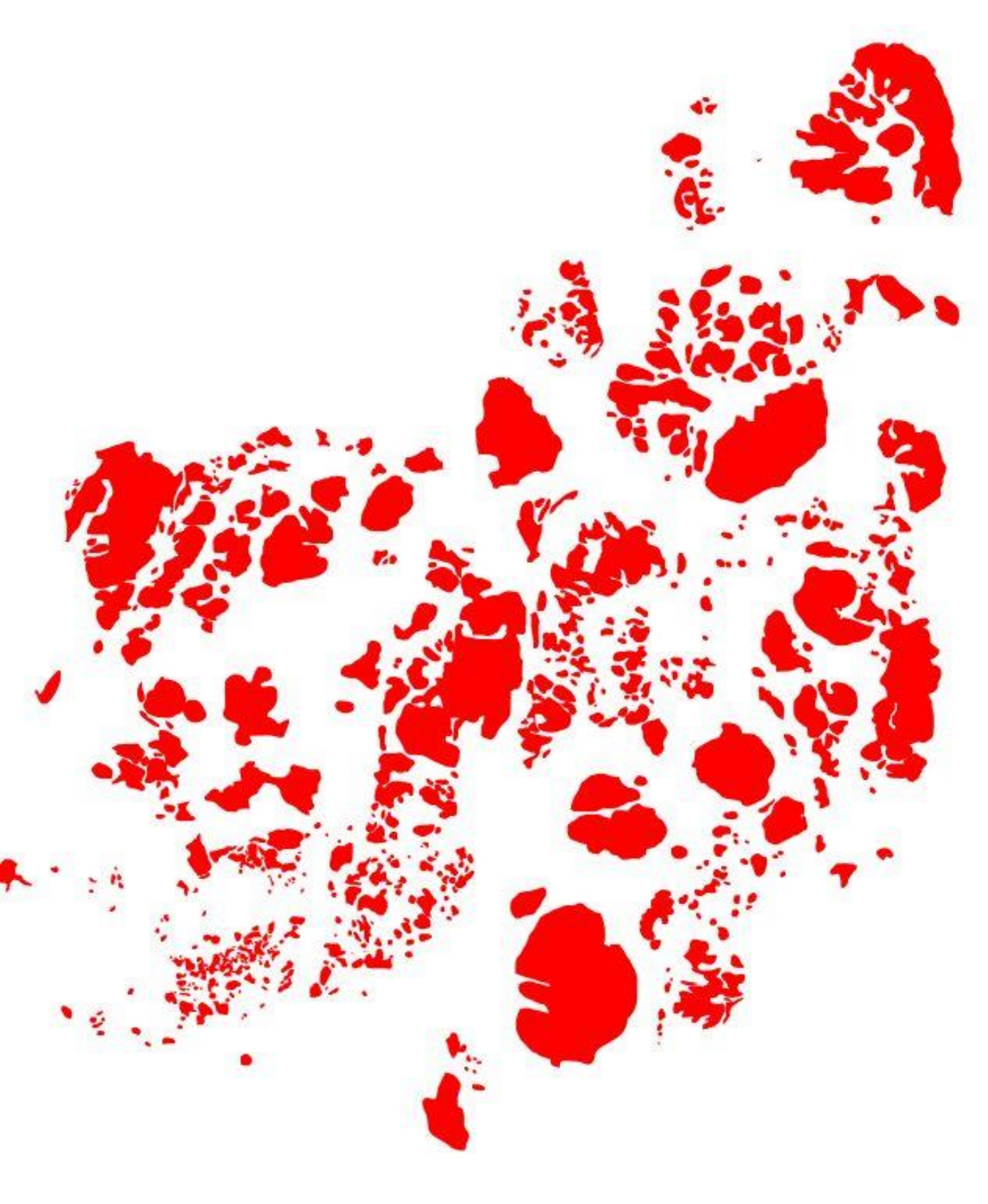,width = 0.32\textwidth}}}
\subfigure[Segmentation]{\frame{\epsfig{figure=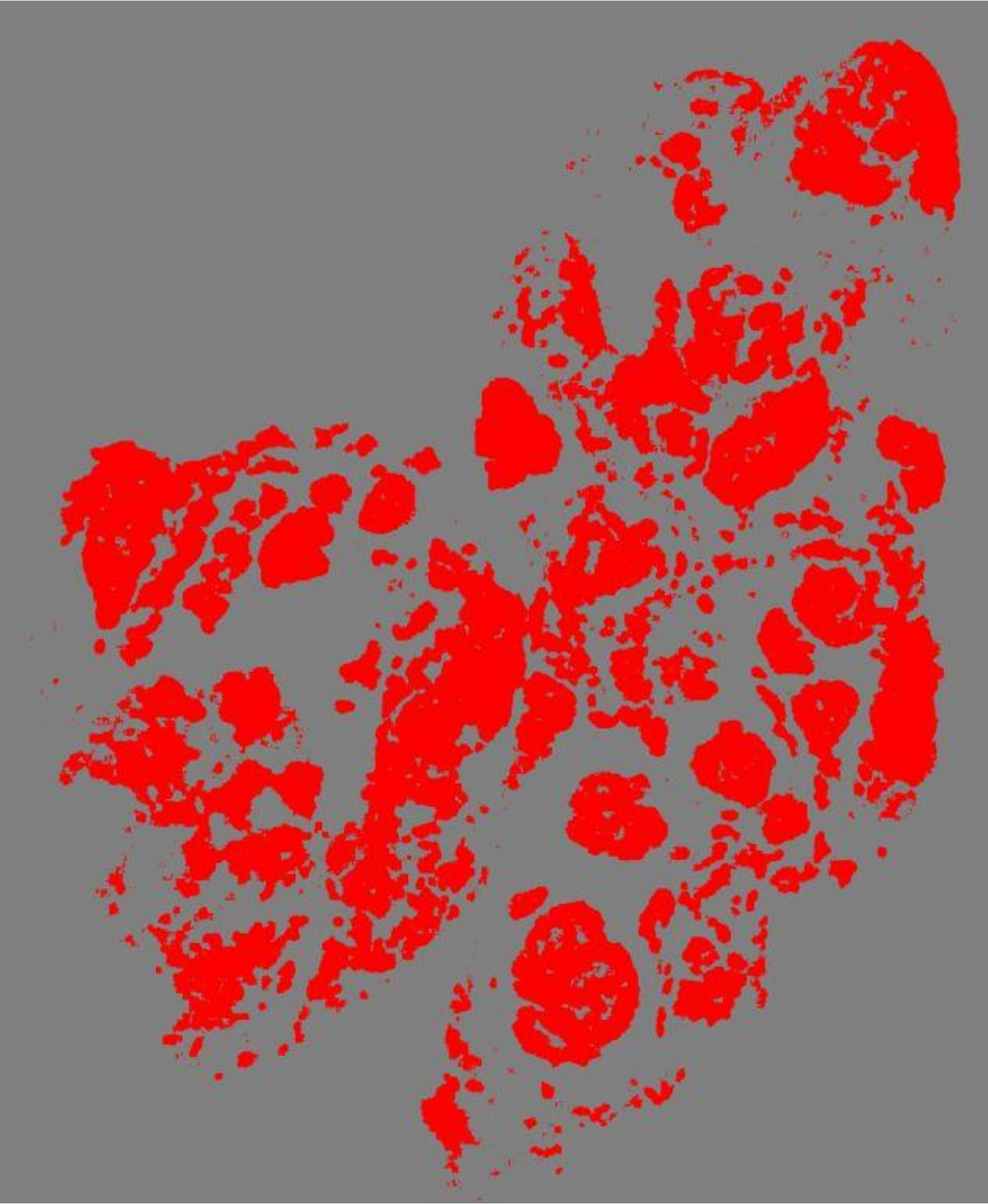,width = 0.32\textwidth}}}

\subfigure[Zoom-in Image]{\frame{\epsfig{figure=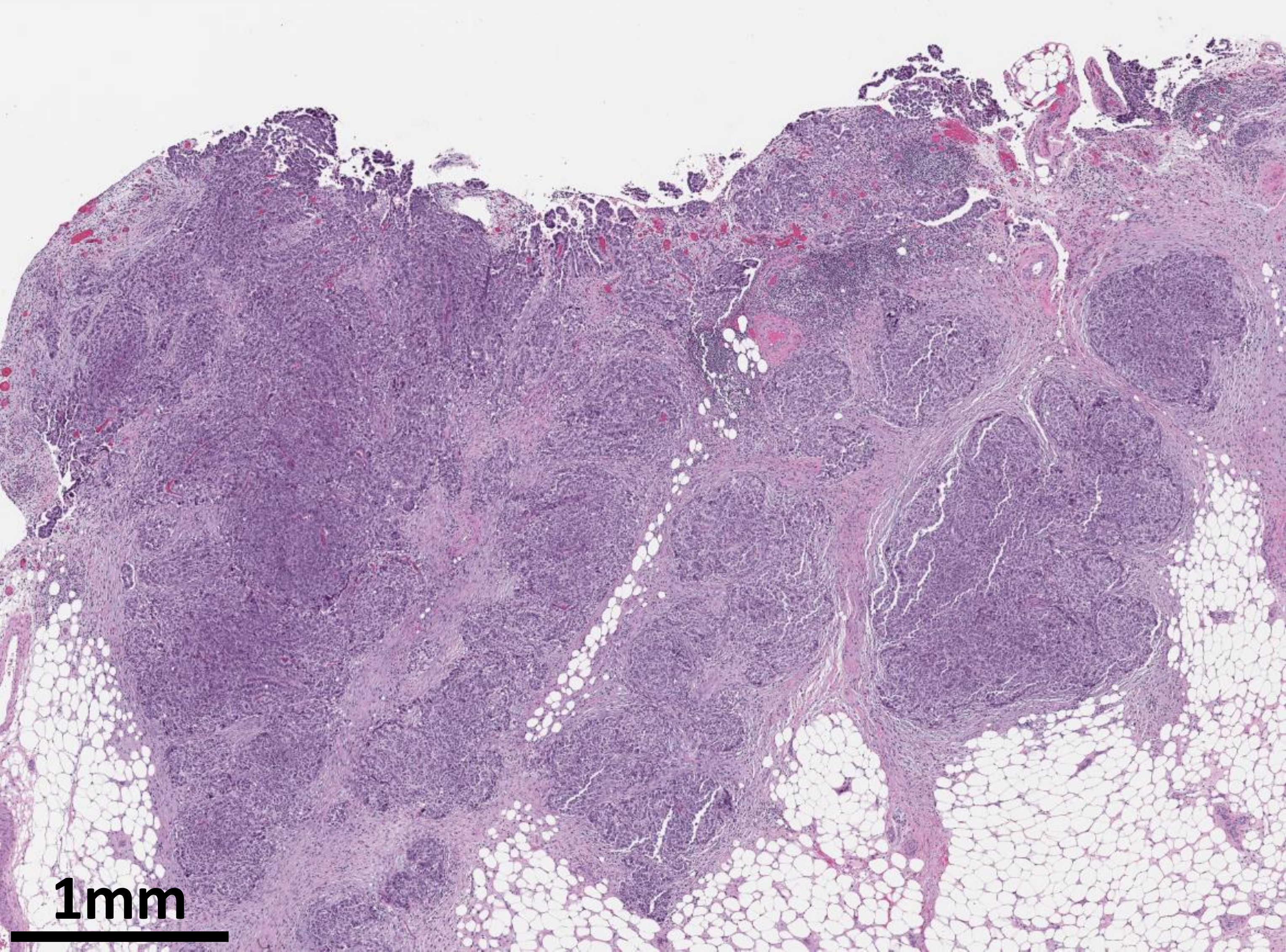,width = 0.32\textwidth}}}
\subfigure[Groundtruth]{\frame{\epsfig{figure=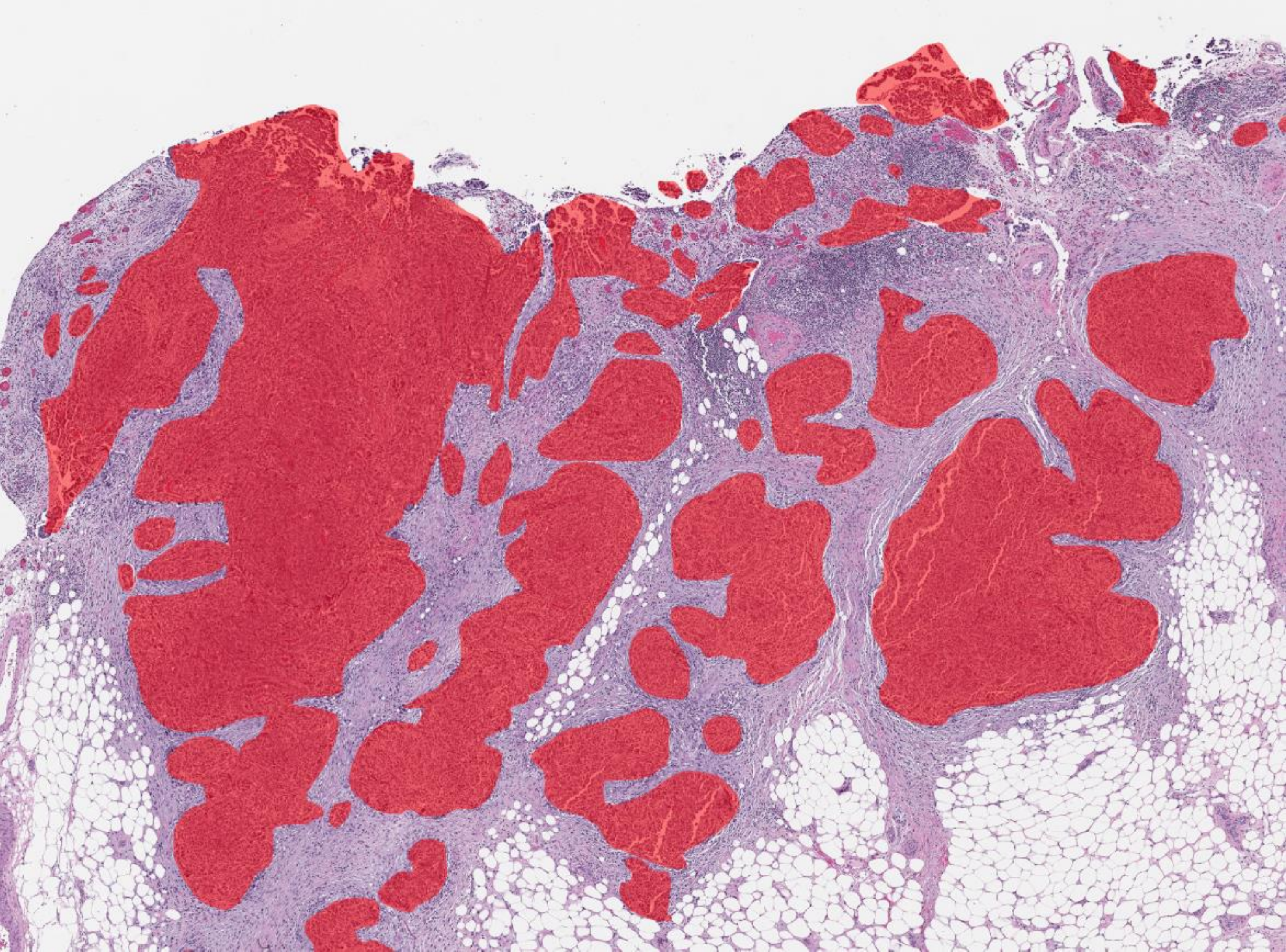,width = 0.32\textwidth}}}
\subfigure[Segmentation]{\frame{\epsfig{figure=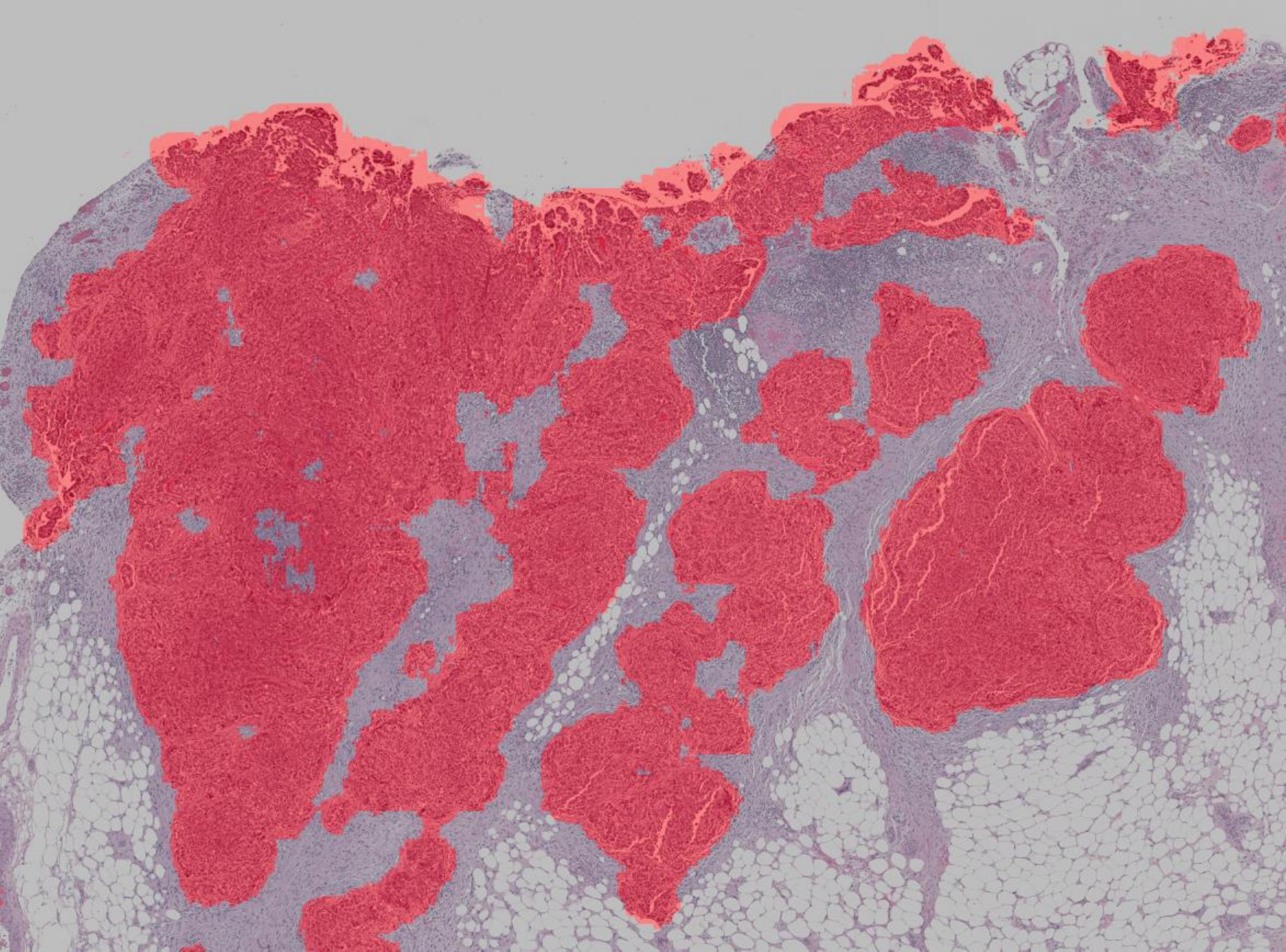,width = 0.32\textwidth}}}
\caption{Ovarian image, its groundtruth, and its segmentation. (a-c) show the entire whole slide image and (d-f) show a zoom-in image. Cancers are highlighted in red. White regions in (b,e) and gray regions in (c,f) are non-cancer.}
\label{fig:img3}
\end{figure*}

\begin{figure*}[ht!]
\centering
\subfigure[Whole Slide Image]{\frame{\epsfig{figure=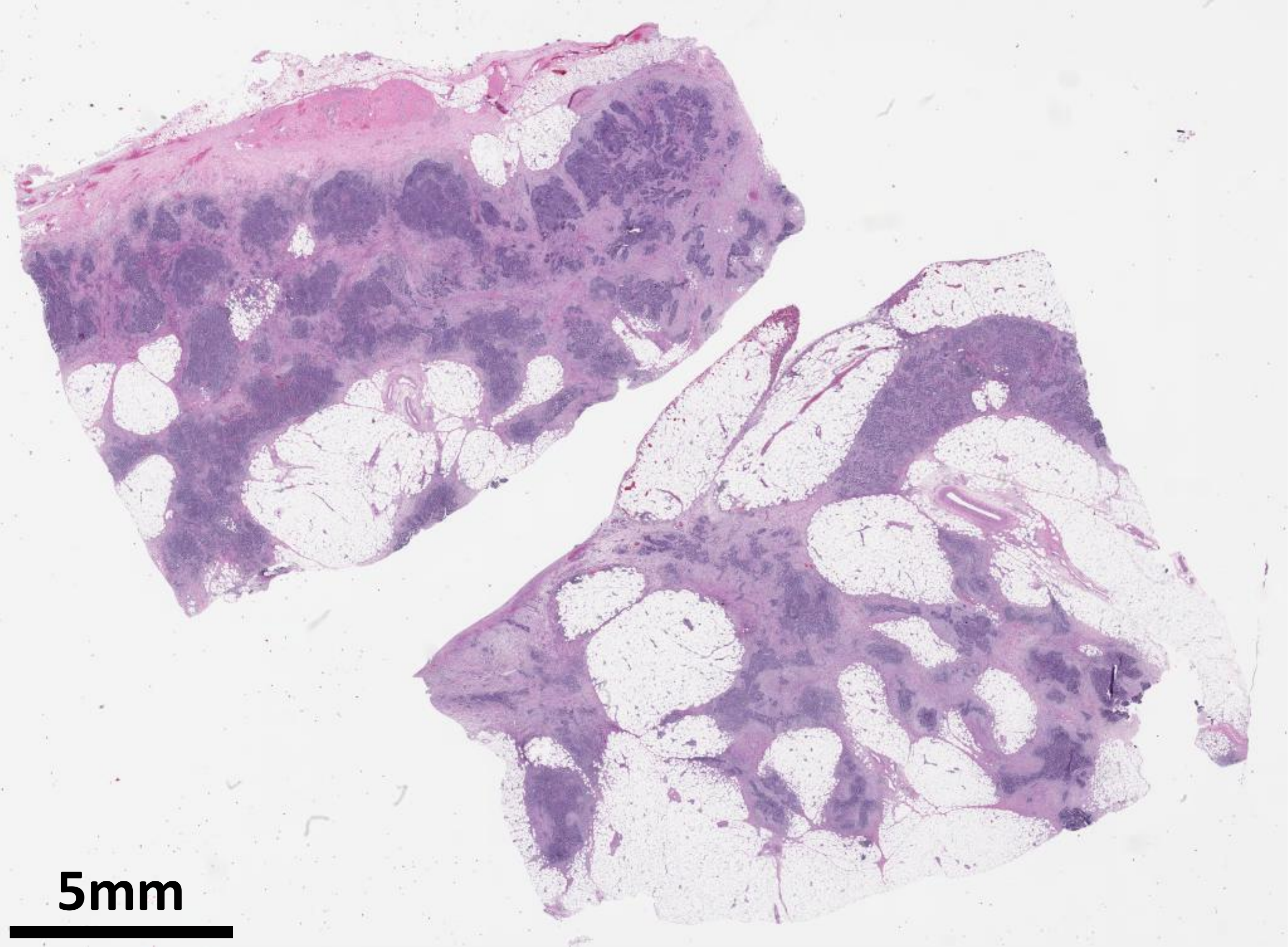,width = 0.32\textwidth}}}
\subfigure[Groundtruth]{\frame{\epsfig{figure=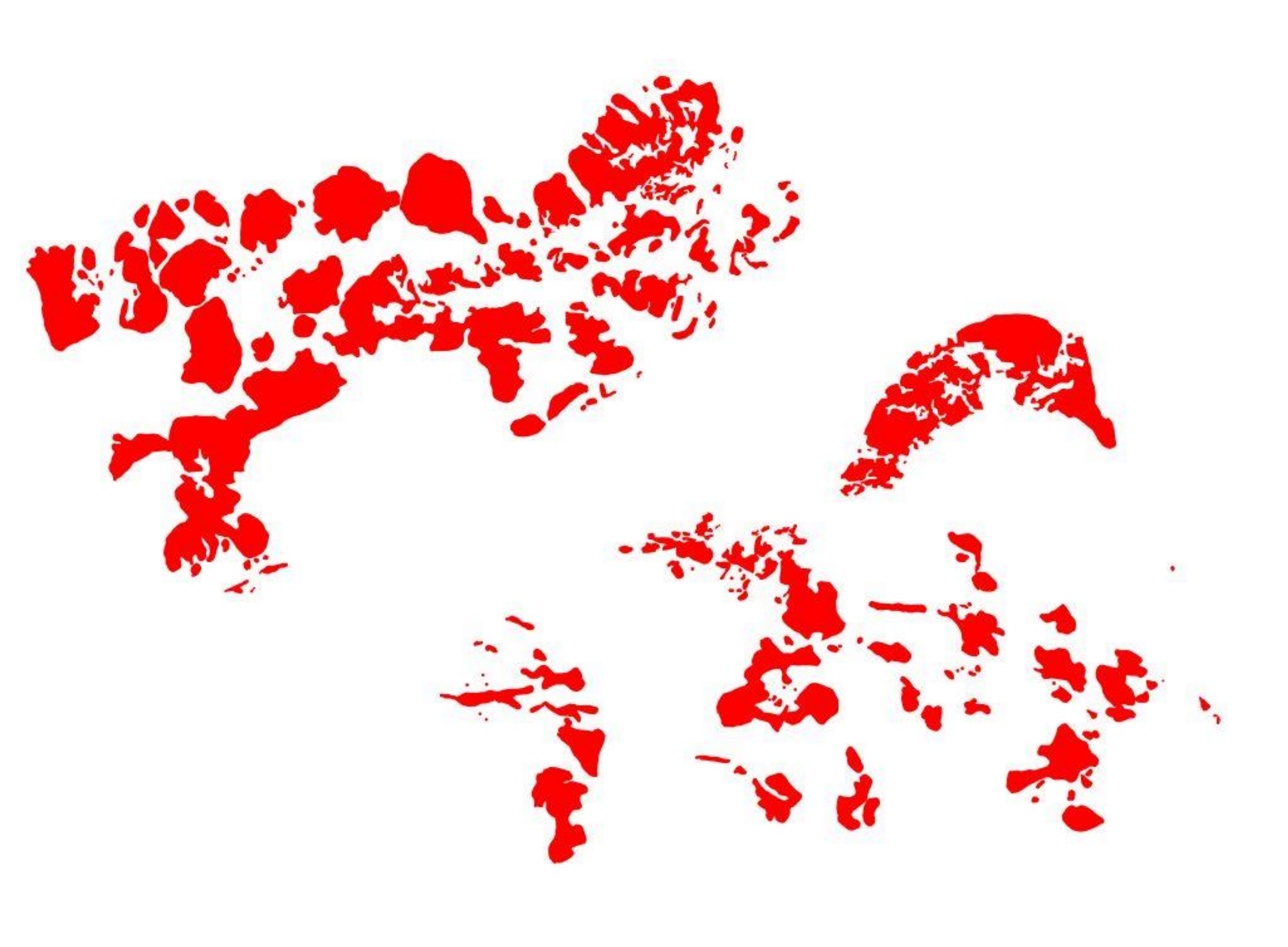,width = 0.32\textwidth}}}
\subfigure[Segmentation]{\frame{\epsfig{figure=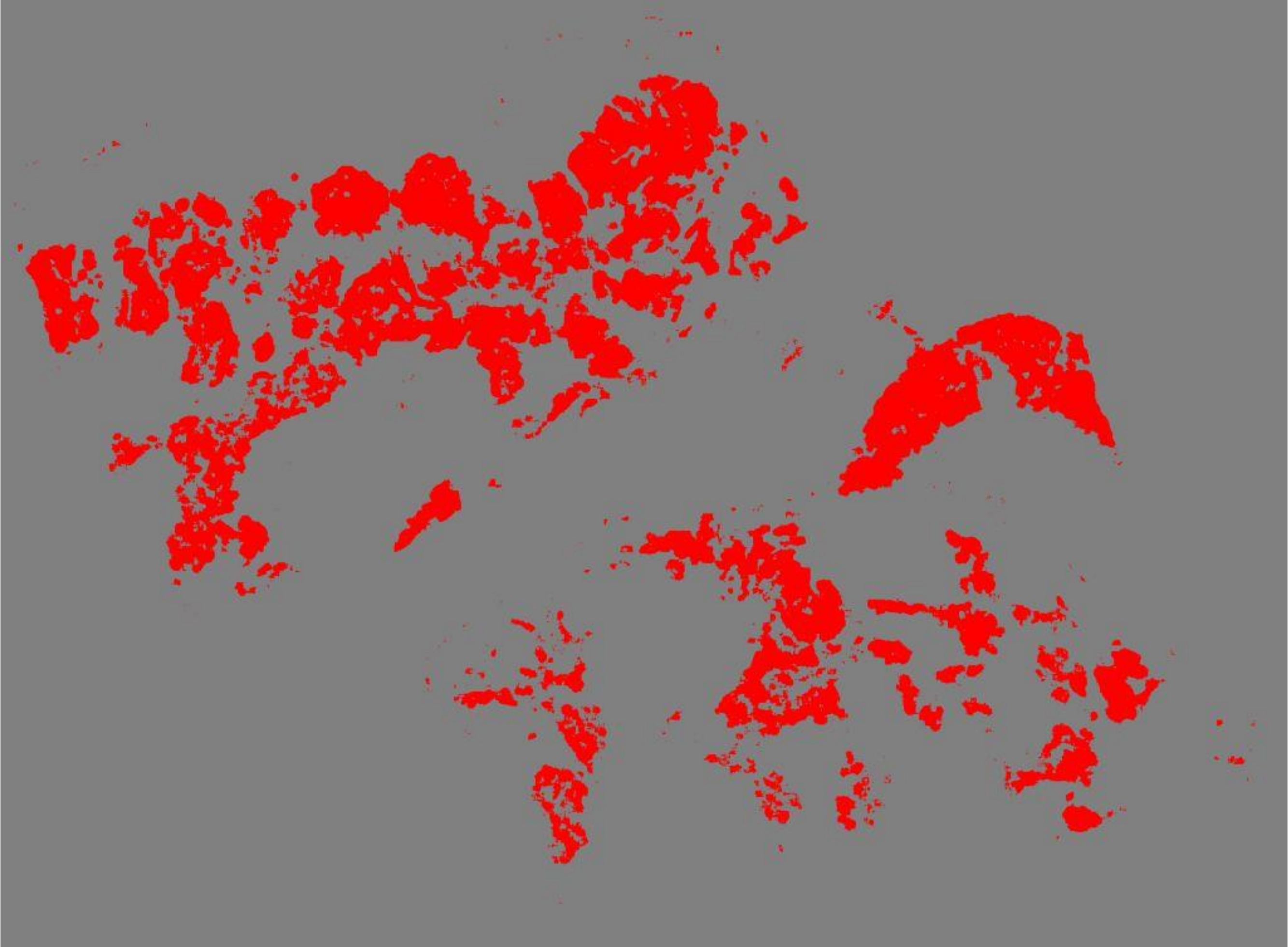,width = 0.32\textwidth}}}

\subfigure[Zoom-in Image]{\frame{\epsfig{figure=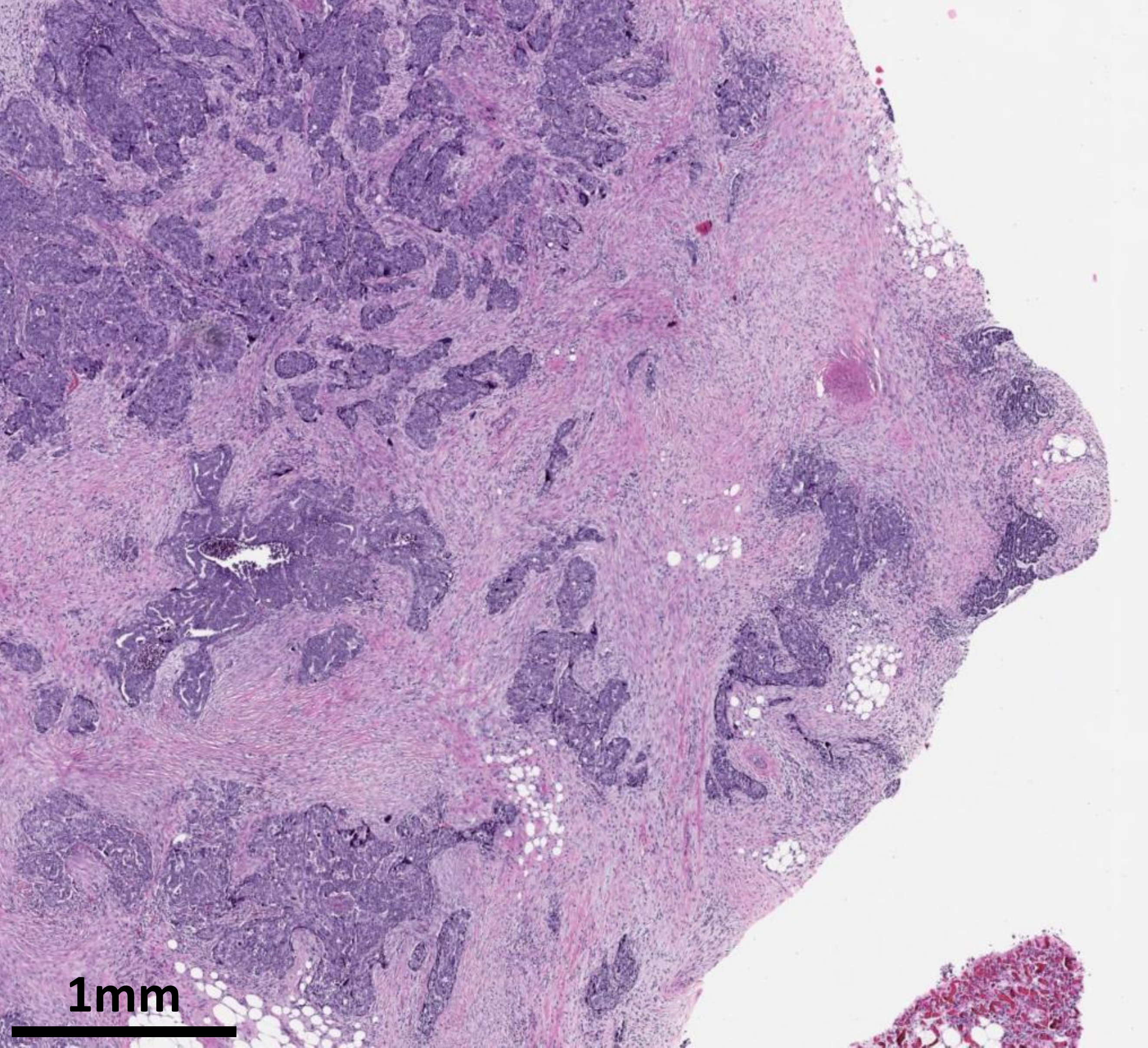,width = 0.32\textwidth}}}
\subfigure[Groundtruth]{\frame{\epsfig{figure=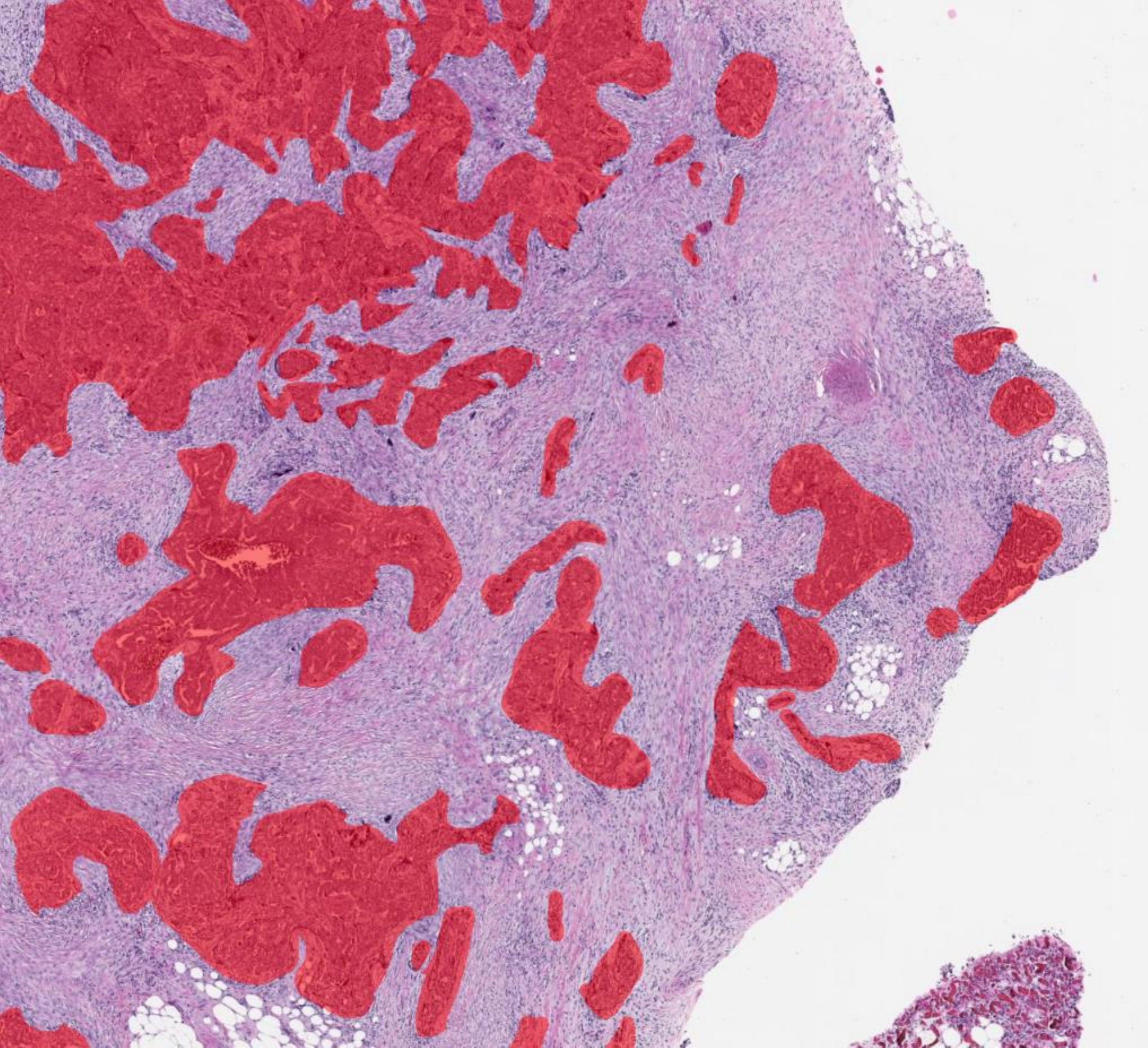,width = 0.32\textwidth}}}
\subfigure[Segmentation]{\frame{\epsfig{figure=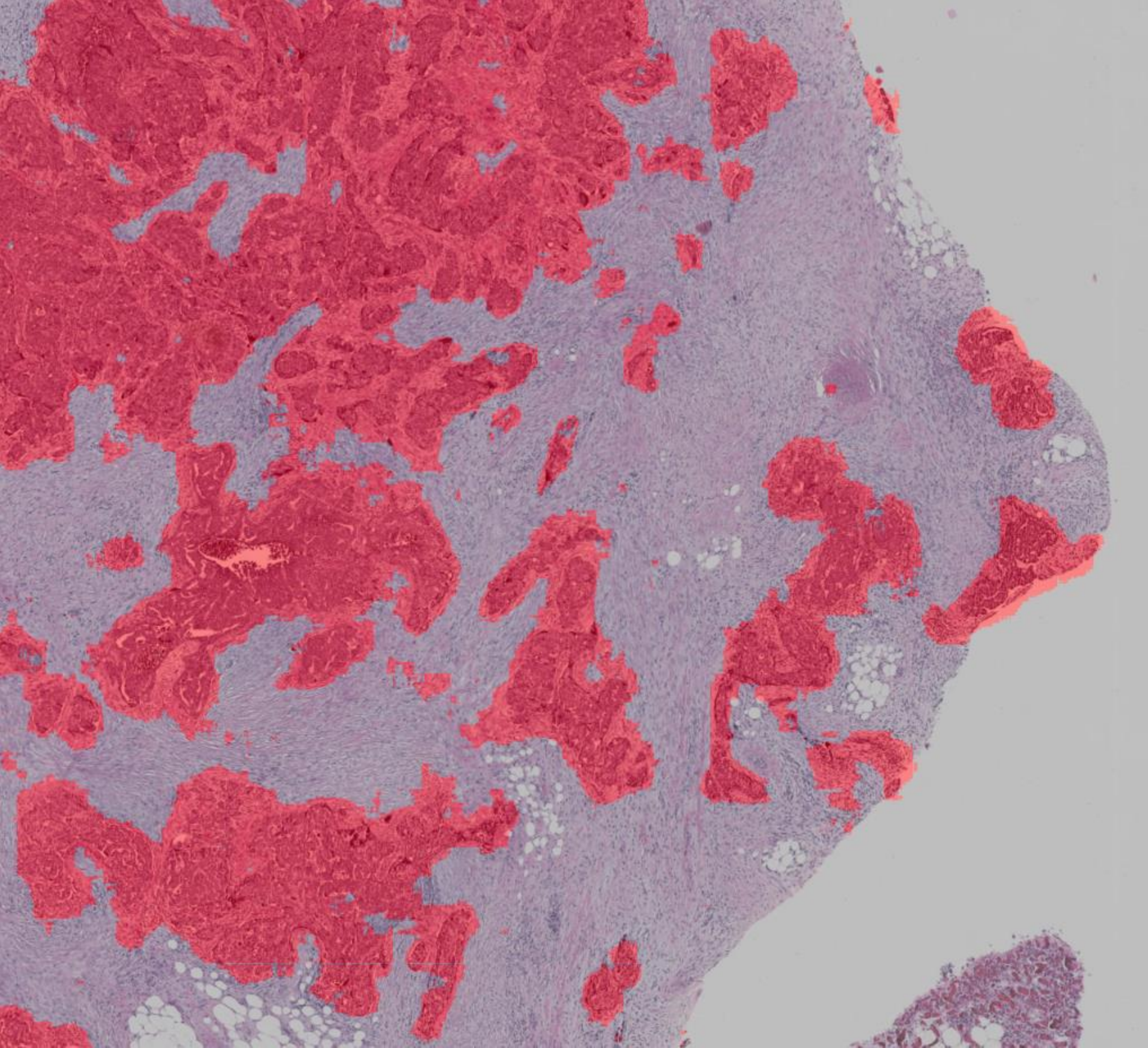,width = 0.32\textwidth}}}
\caption{Ovarian image, its groundtruth, and its segmentation. (a-c) show the entire whole slide image and (d-f) show a zoom-in image. Cancers are highlighted in red. White regions in (b,e) and gray regions in (c,f) are non-cancer.}
\label{fig:img2}
\end{figure*}

\begin{figure*}[ht!]
\centering
\subfigure[Whole Slide Image]{\frame{\epsfig{figure=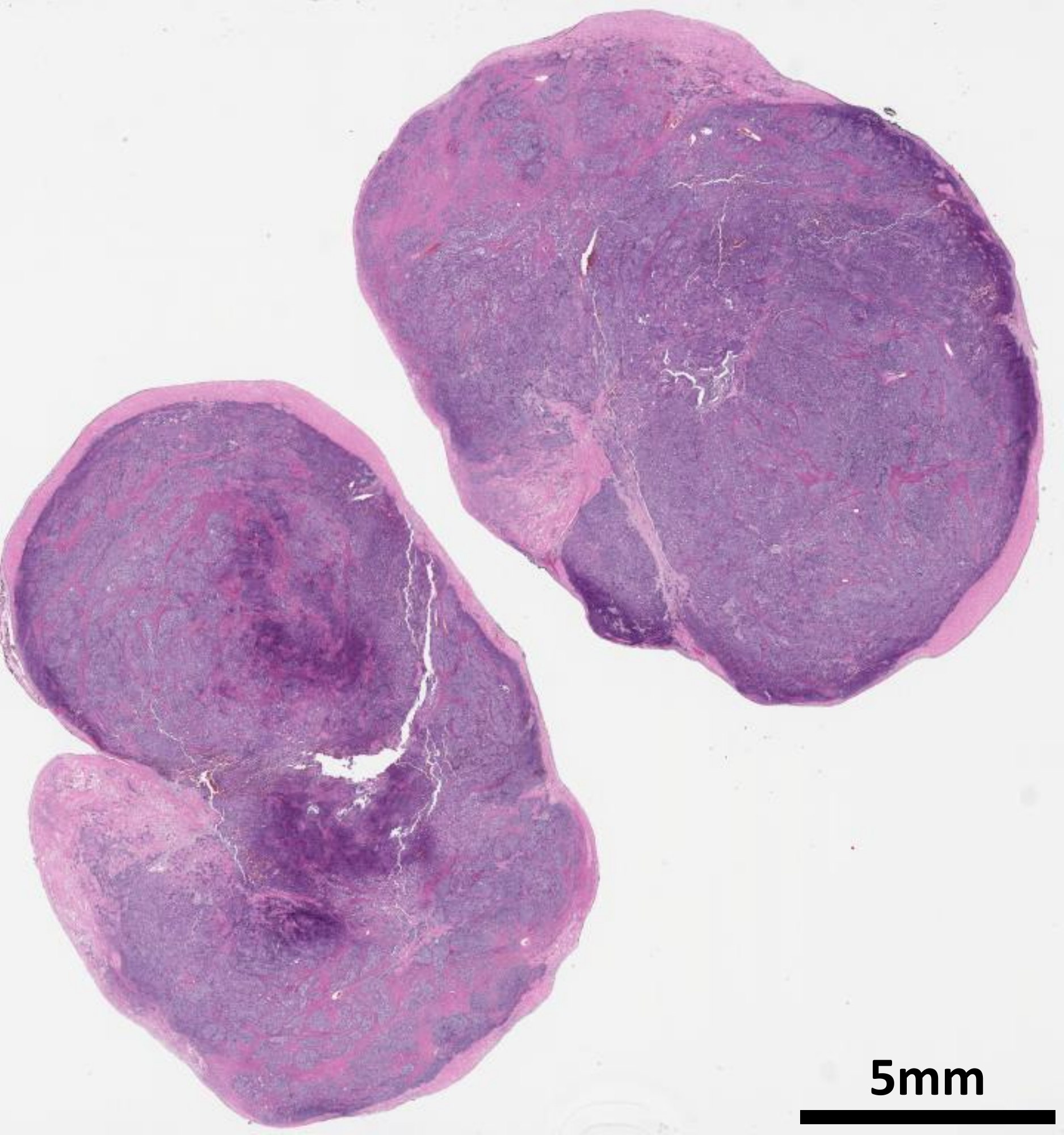,width = 0.32\textwidth}}}
\subfigure[Groundtruth]{\frame{\epsfig{figure=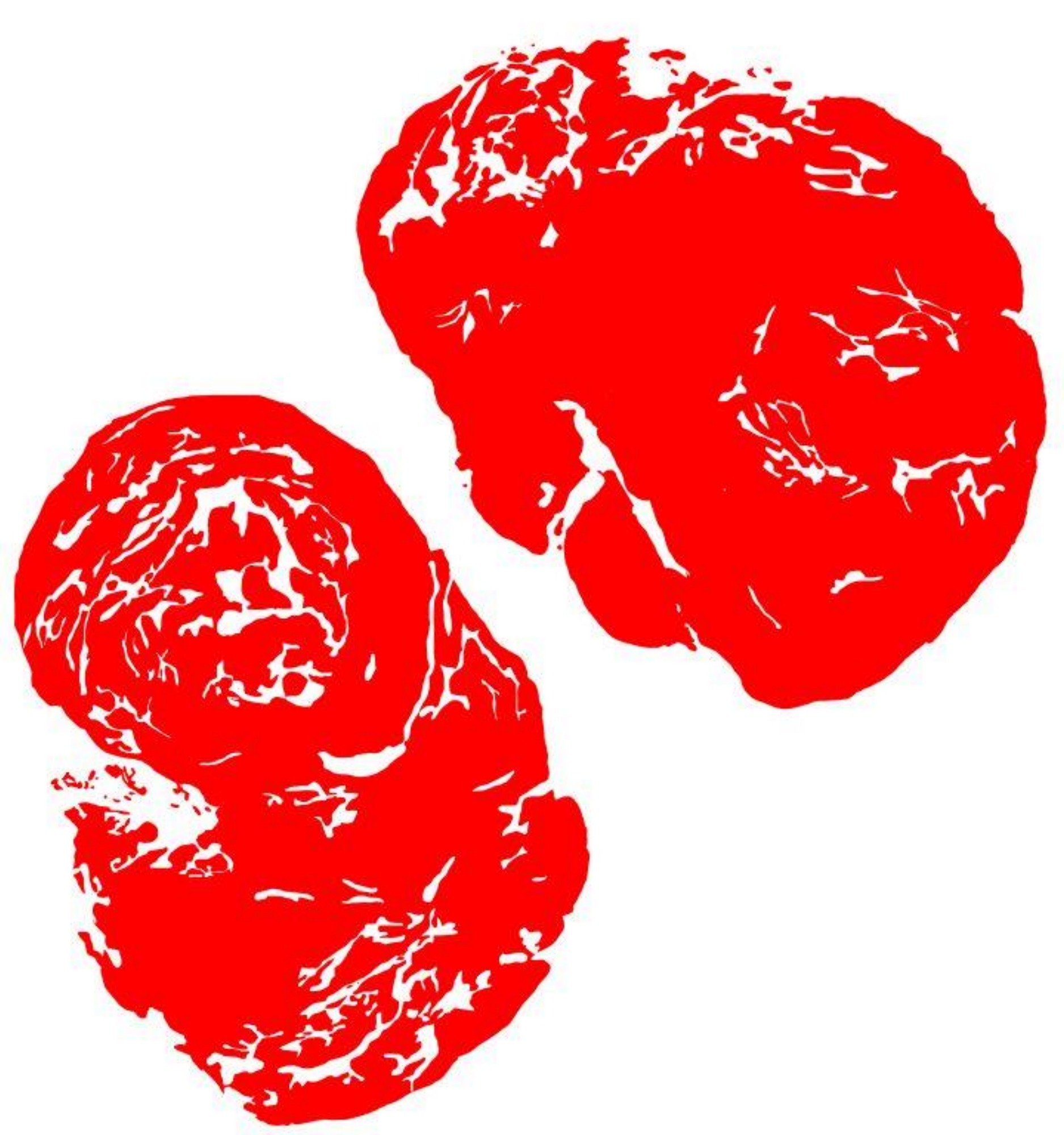,width = 0.32\textwidth}}}
\subfigure[Segmentation]{\frame{\epsfig{figure=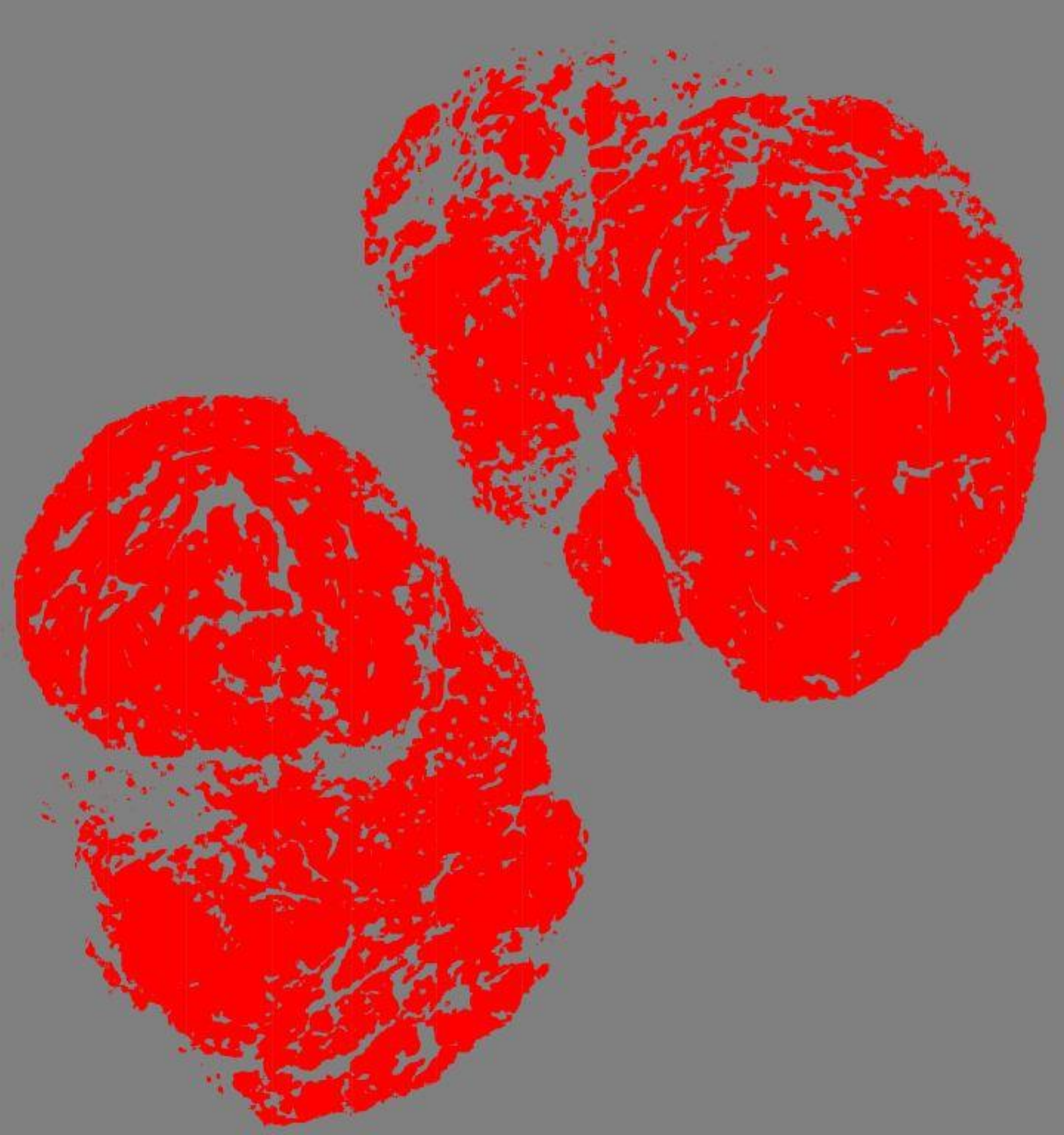,width = 0.32\textwidth}}}

\subfigure[Zoom-in Image]{\frame{\epsfig{figure=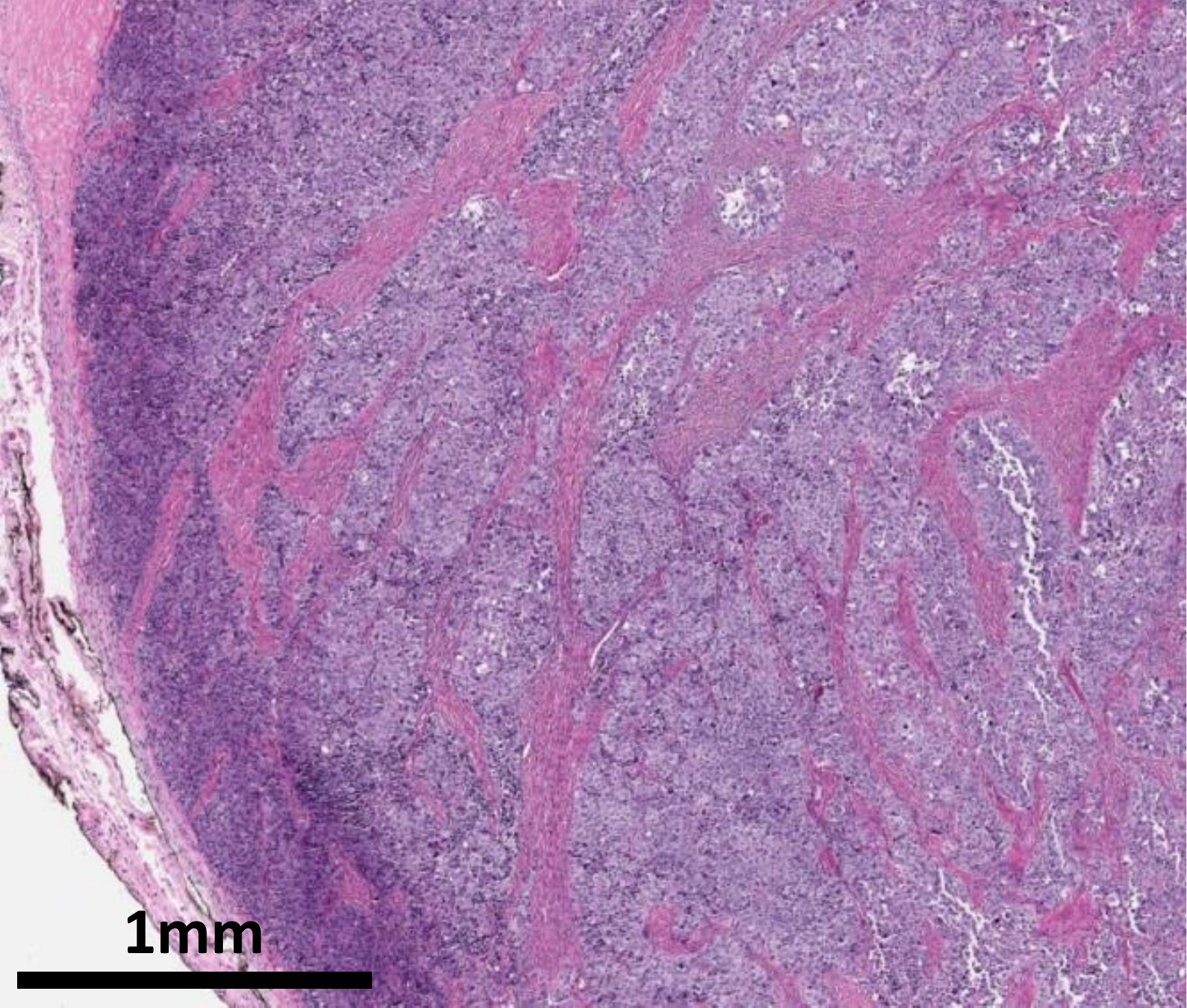,width = 0.32\textwidth}}}
\subfigure[Groundtruth]{\frame{\epsfig{figure=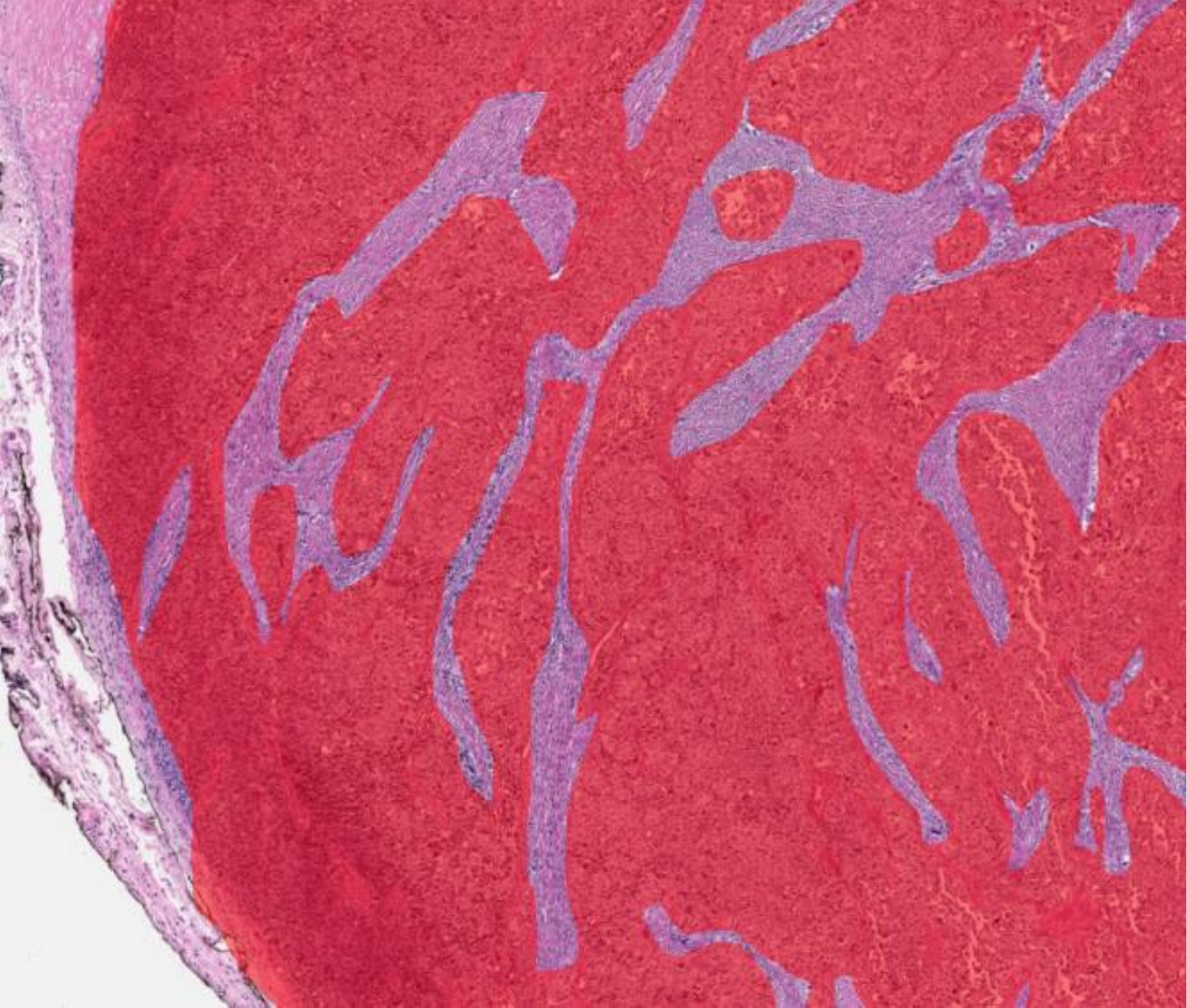,width = 0.32\textwidth}}}
\subfigure[Segmentation]{\frame{\epsfig{figure=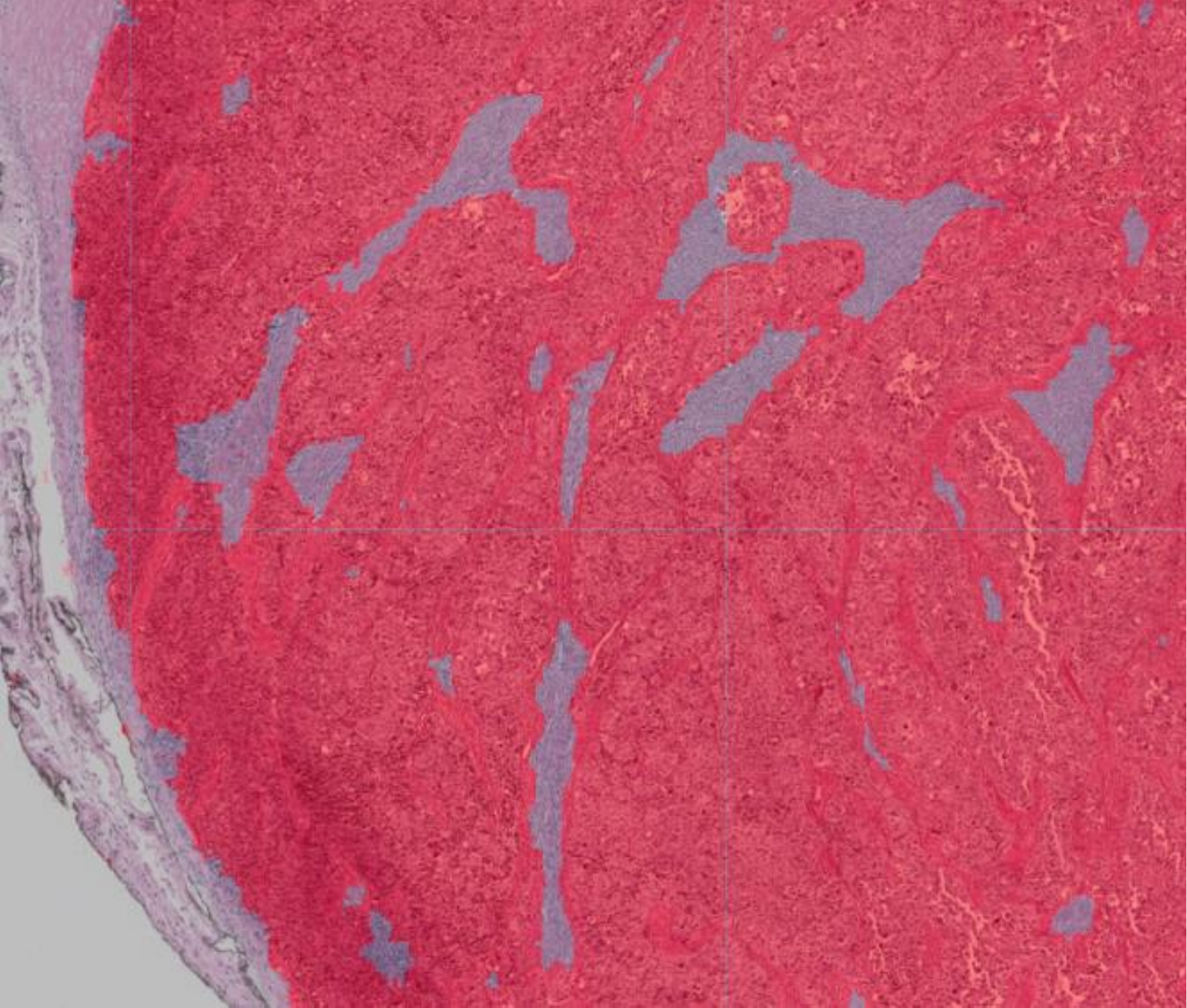,width = 0.32\textwidth}}}
\caption{Ovarian image, its groundtruth, and its segmentation. (a-c) show the entire whole slide image and (d-f) show a zoom-in image. Cancers are highlighted in red. White regions in (b,e) and gray regions in (c,f) are non-cancer.}
\label{fig:img1}
\end{figure*}

\begin{figure*}[ht!]
\centering
\subfigure[Image]{\frame{\epsfig{figure=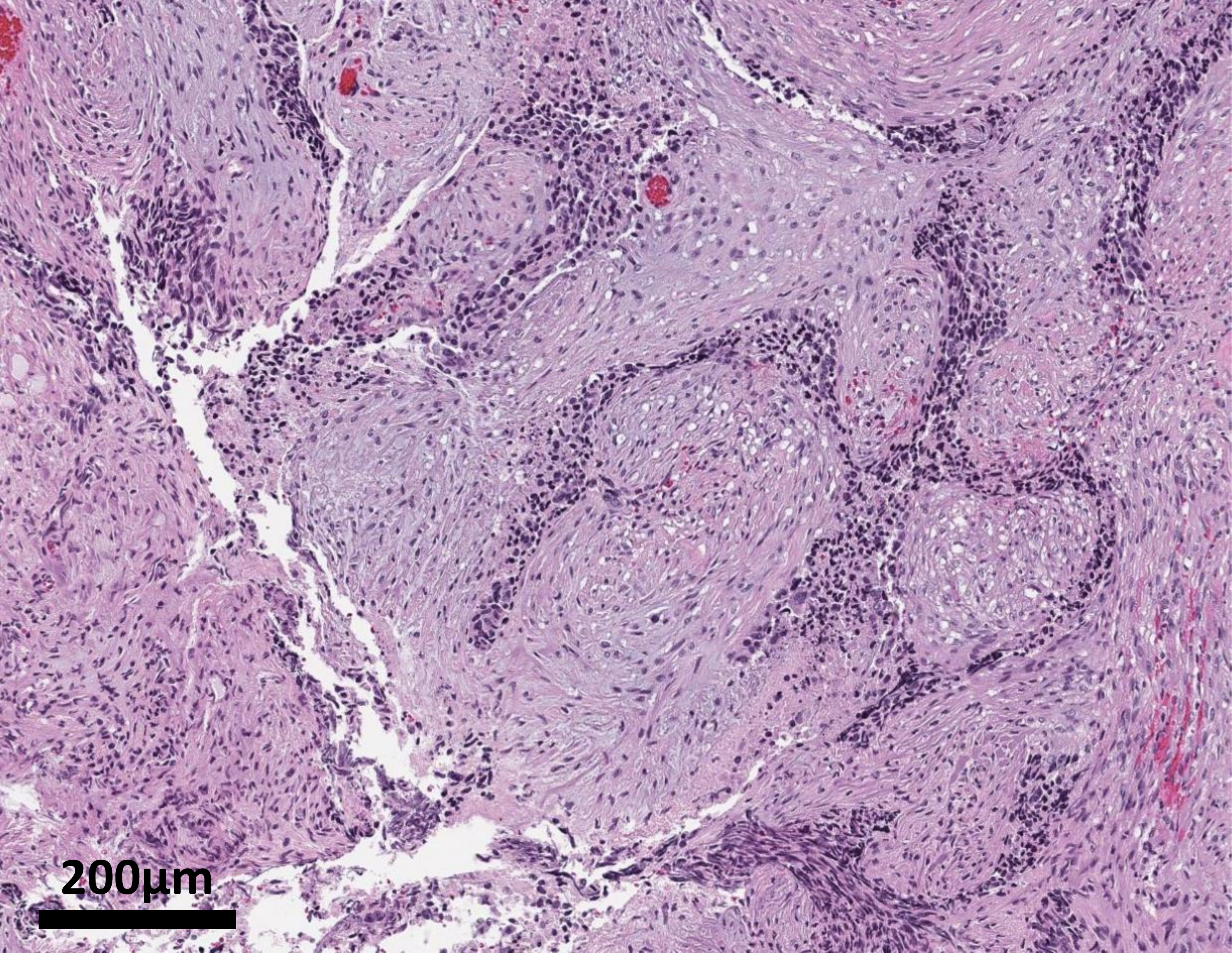,width = 0.48\textwidth}}}
\subfigure[Segmentation]{\frame{\epsfig{figure=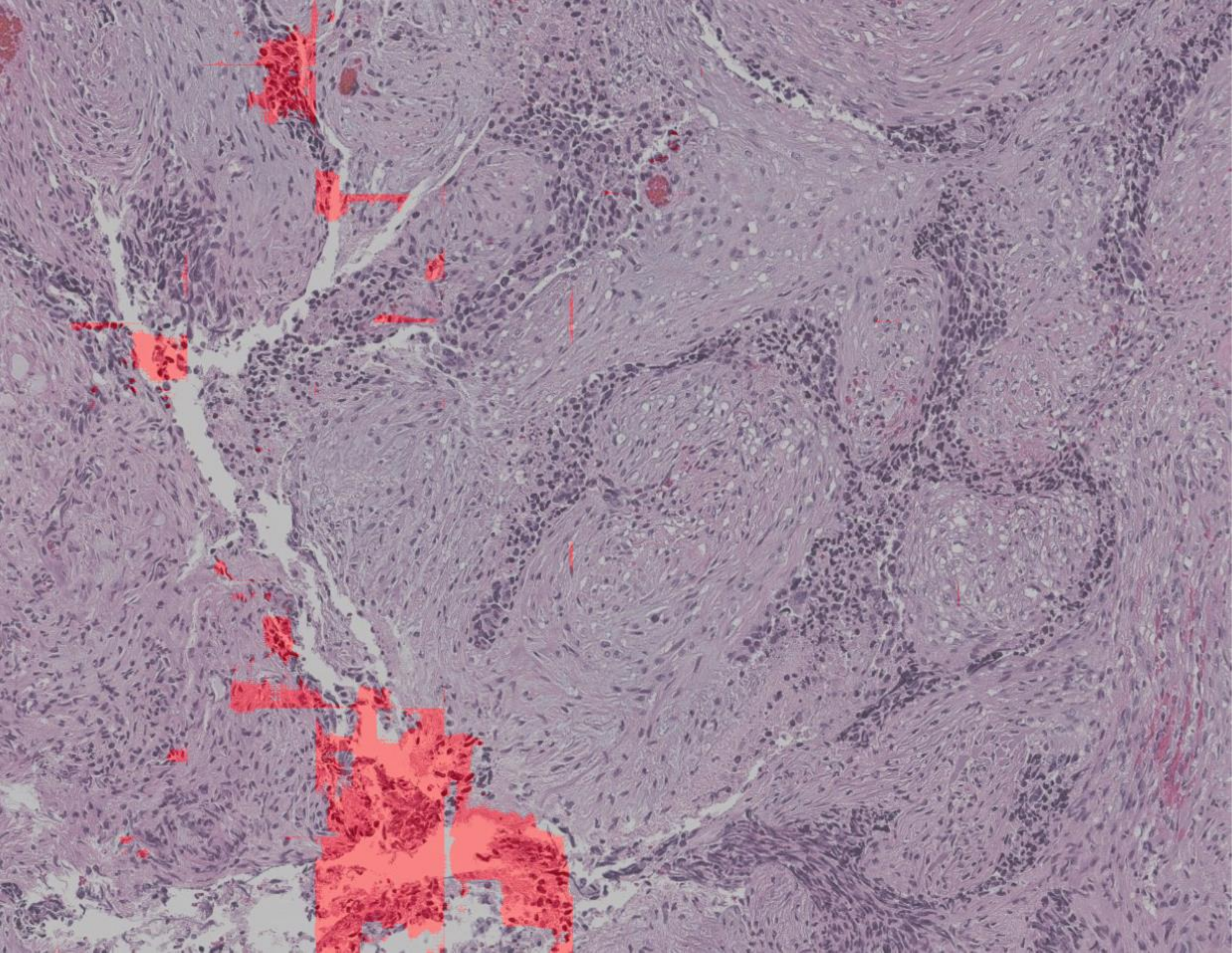,width = 0.48\textwidth}}}

\subfigure[Image]{\frame{\epsfig{figure=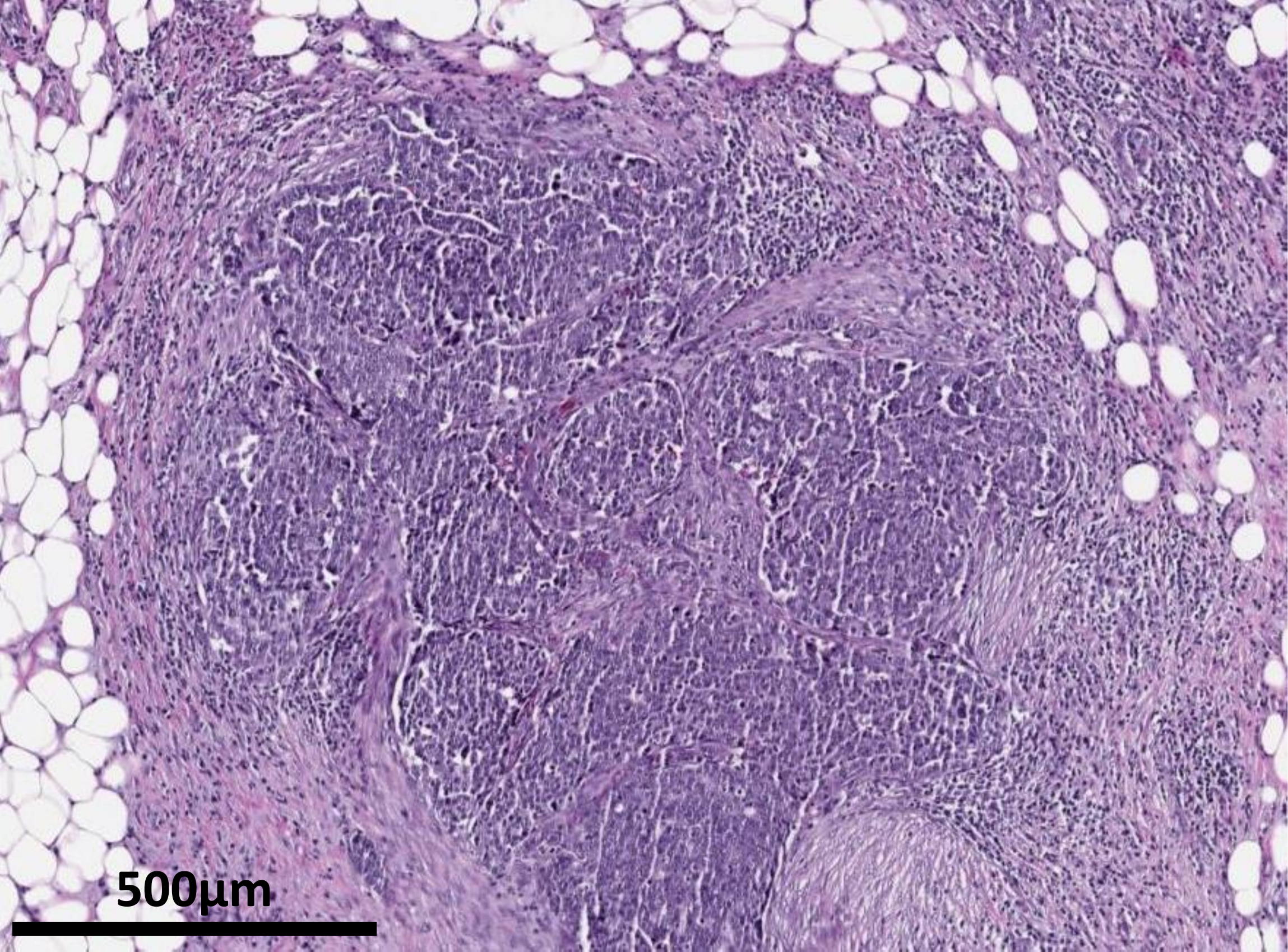,width = 0.48\textwidth}}}
\subfigure[Segmentation]{\frame{\epsfig{figure=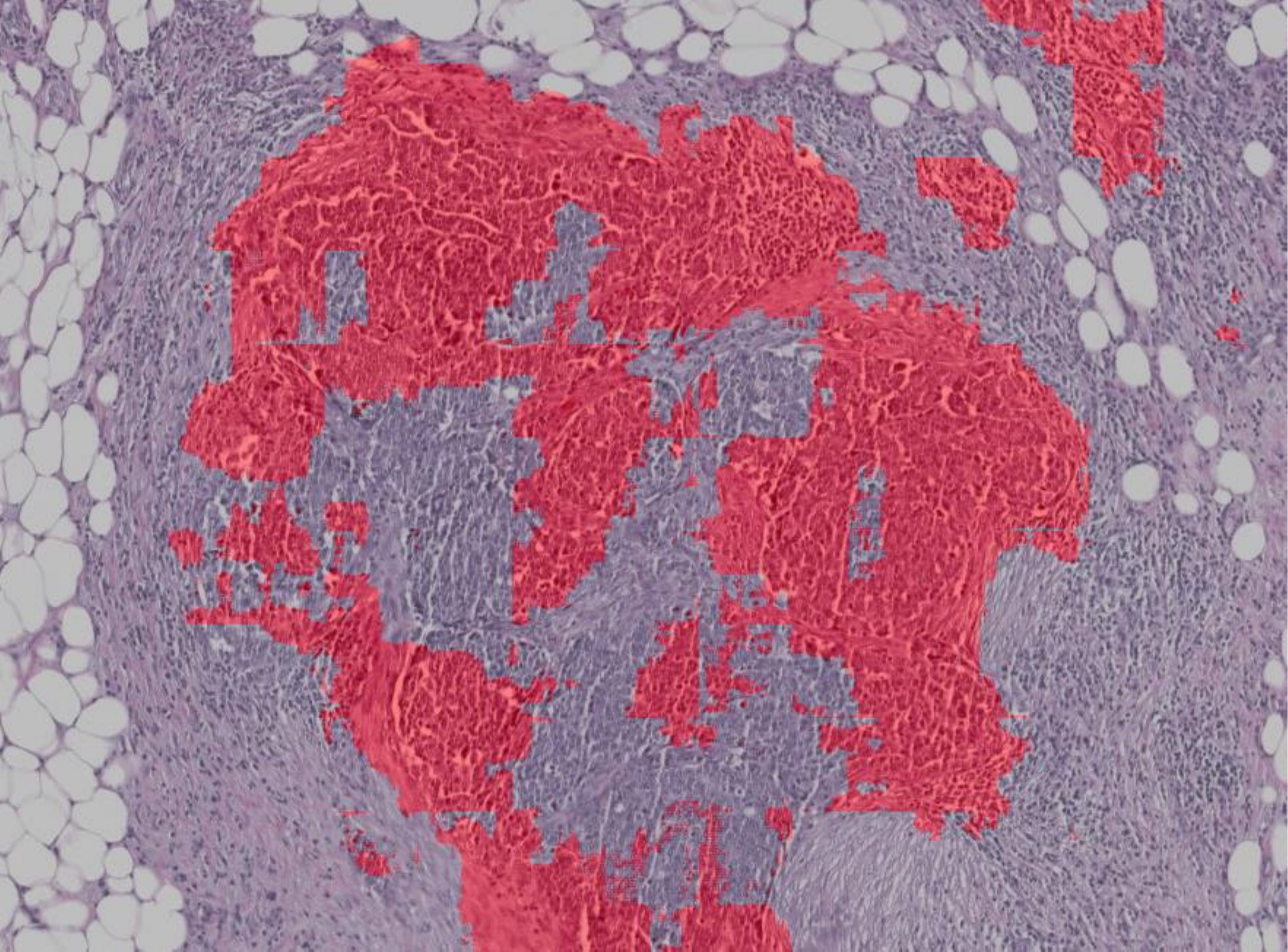,width = 0.48\textwidth}}}
\caption{False negatives of our cancer segmentation model. (a,b) Cautery artifact. (c,d) Poor staining. Red regions indicate cancer predicted by the segmentation model.}
\label{fig:FN}
\end{figure*}

\begin{figure*}[ht!]
\centering
\subfigure[Image]{\frame{\epsfig{figure=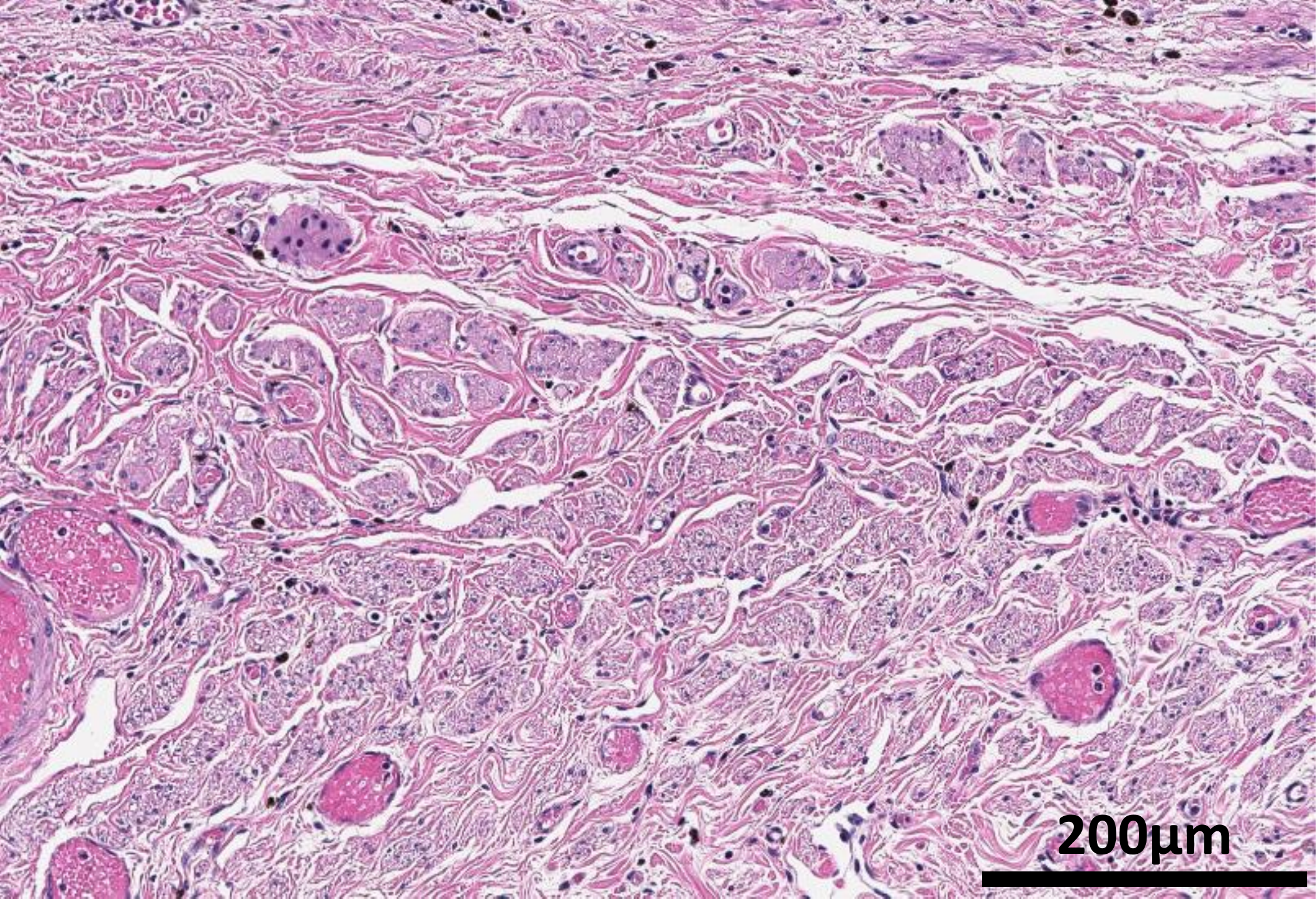,width = 0.47\textwidth}}}
\subfigure[Segmentation]{\frame{\epsfig{figure=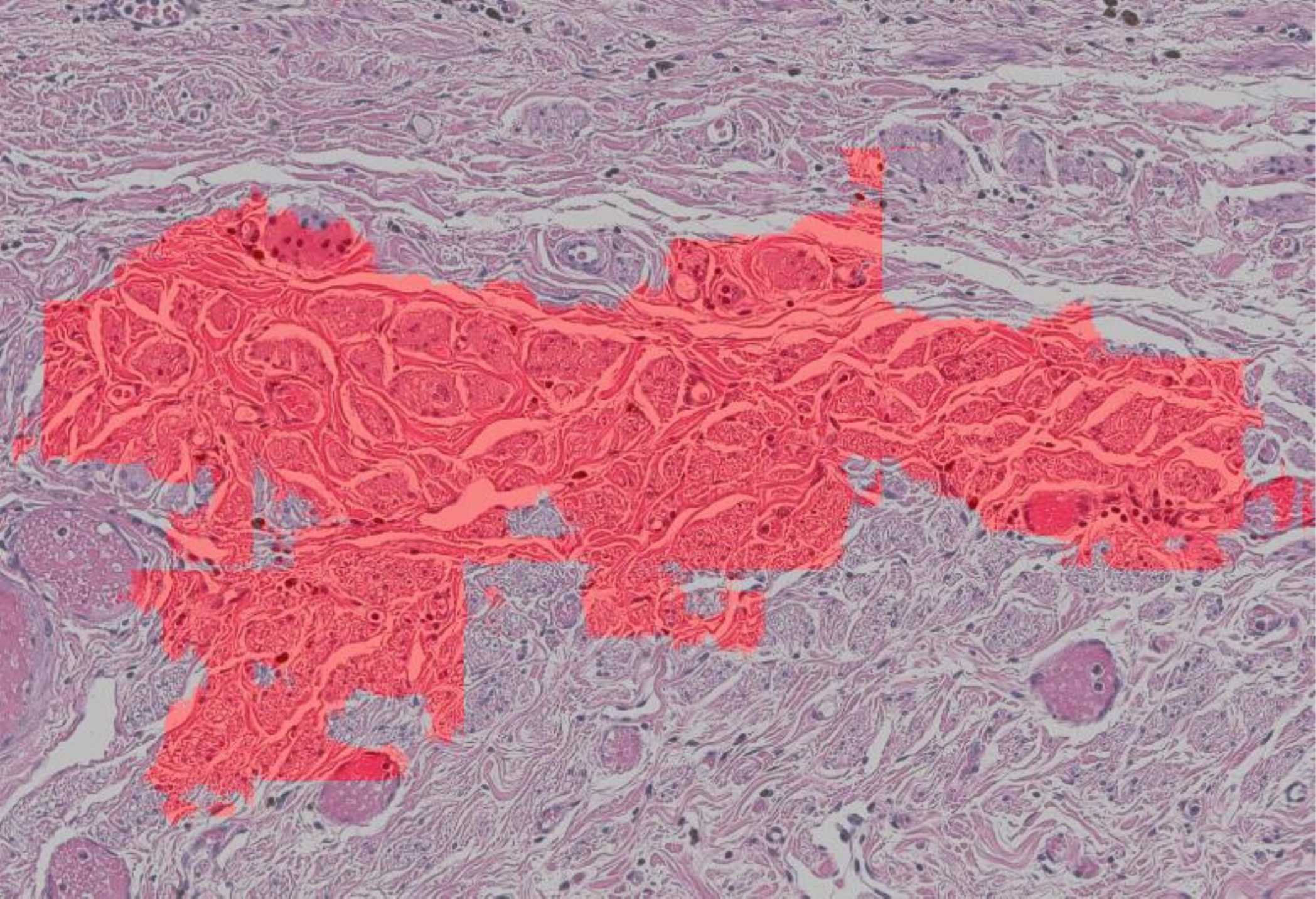,width = 0.47\textwidth}}}

\subfigure[Image]{\frame{\epsfig{figure=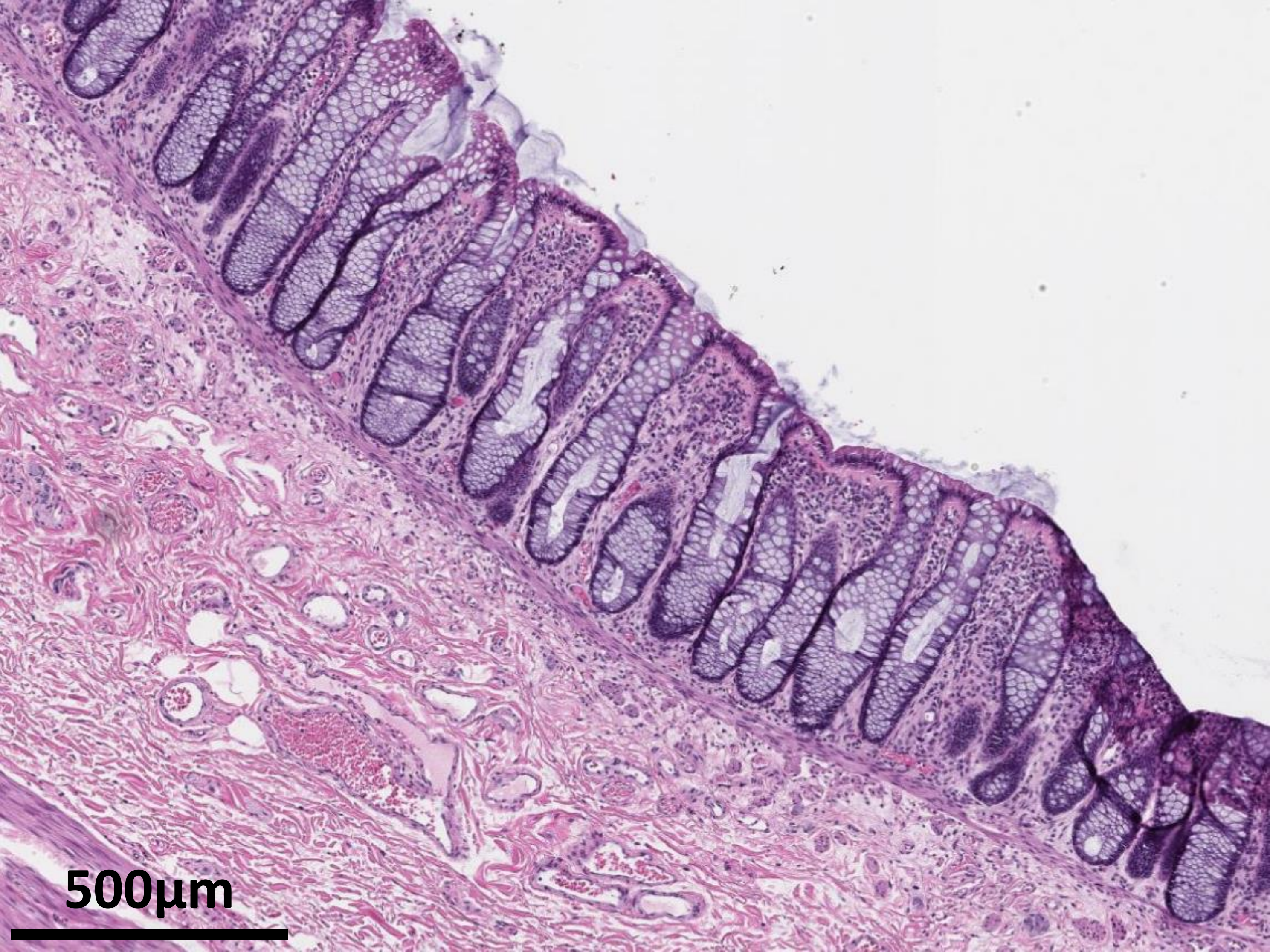,width = 0.47\textwidth}}}
\subfigure[Segmentation]{\frame{\epsfig{figure=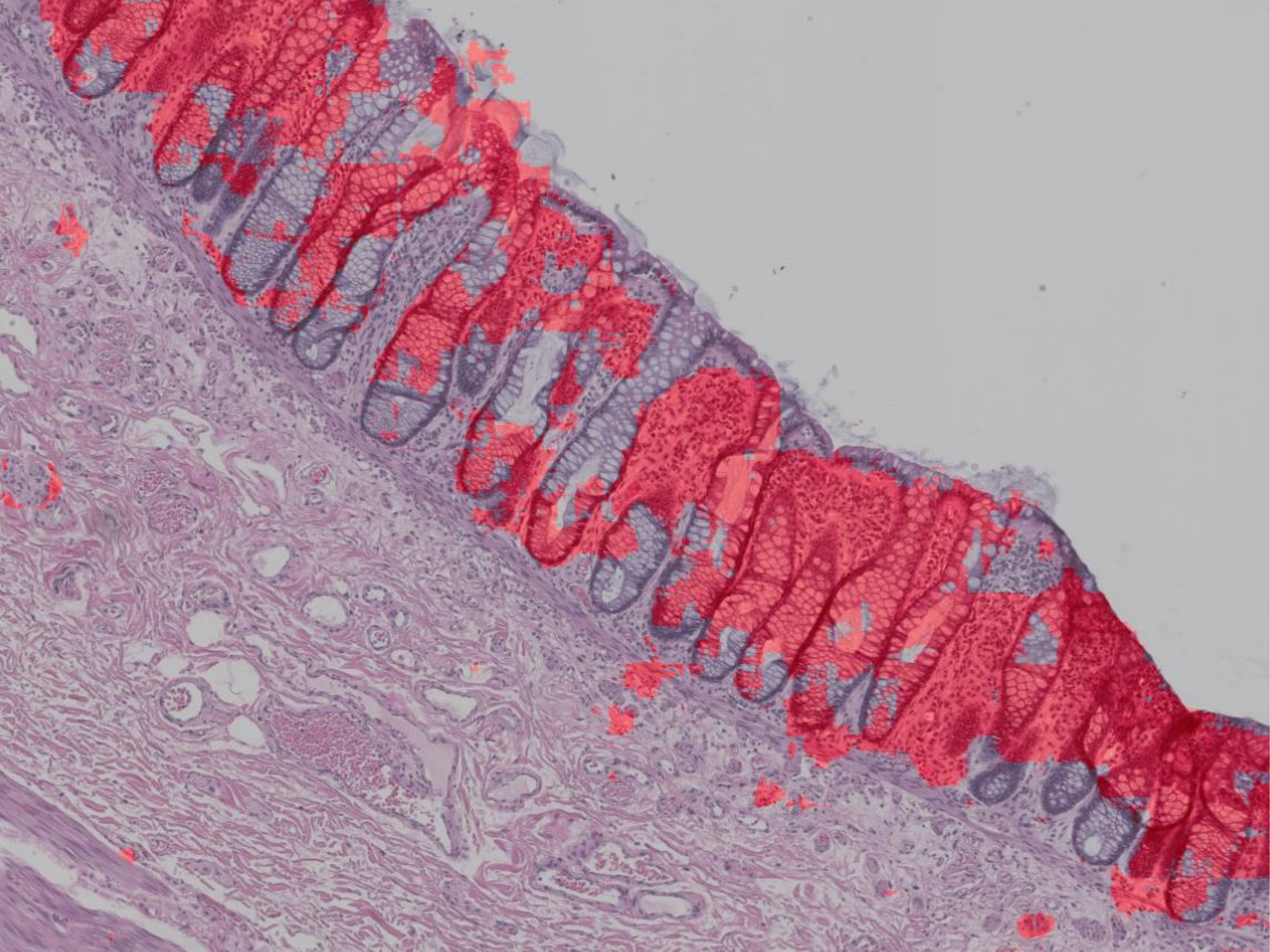,width = 0.47\textwidth}}}

\subfigure[Image]{\frame{\epsfig{figure=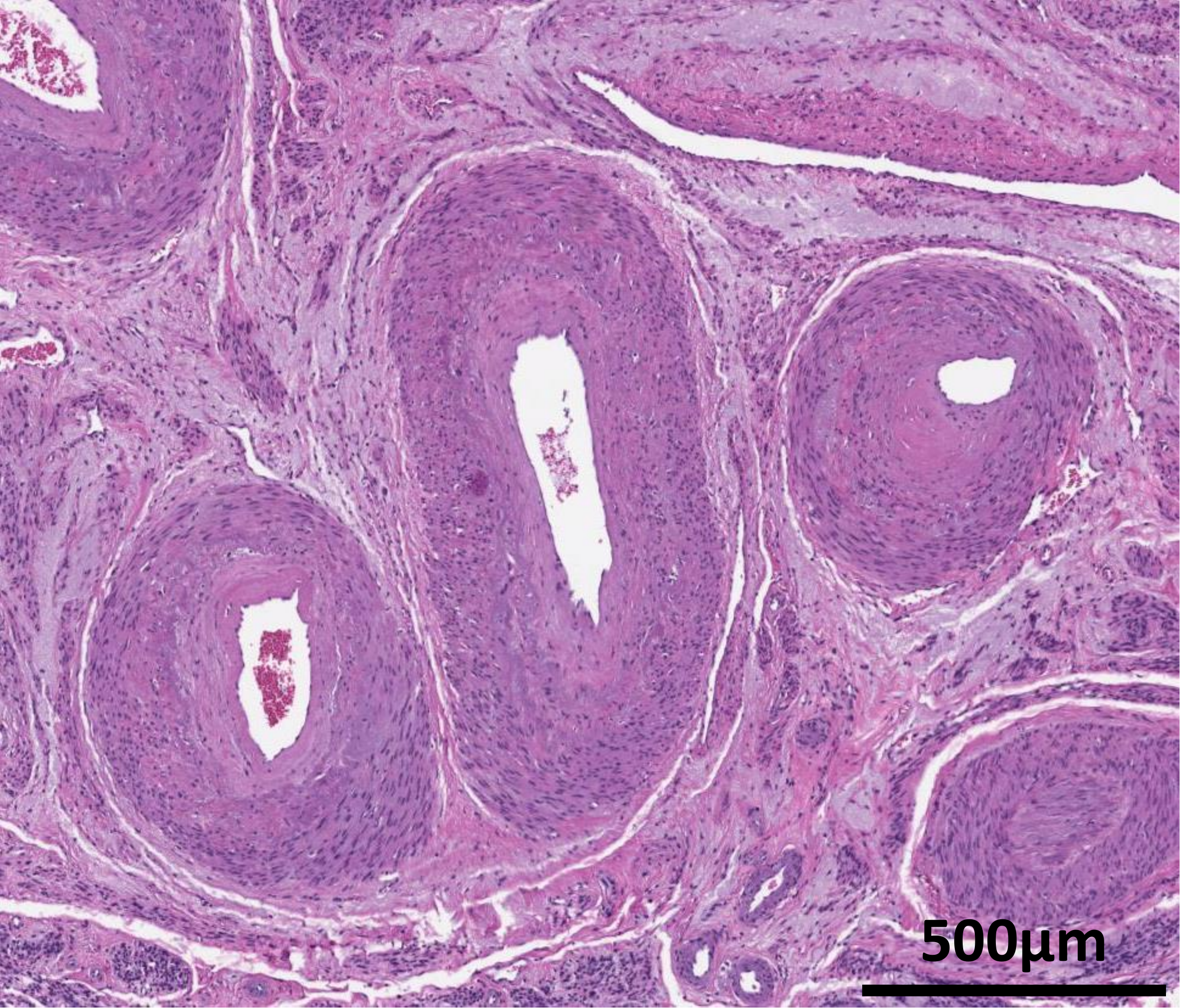,width = 0.47\textwidth}}}
\subfigure[Segmentation]{\frame{\epsfig{figure=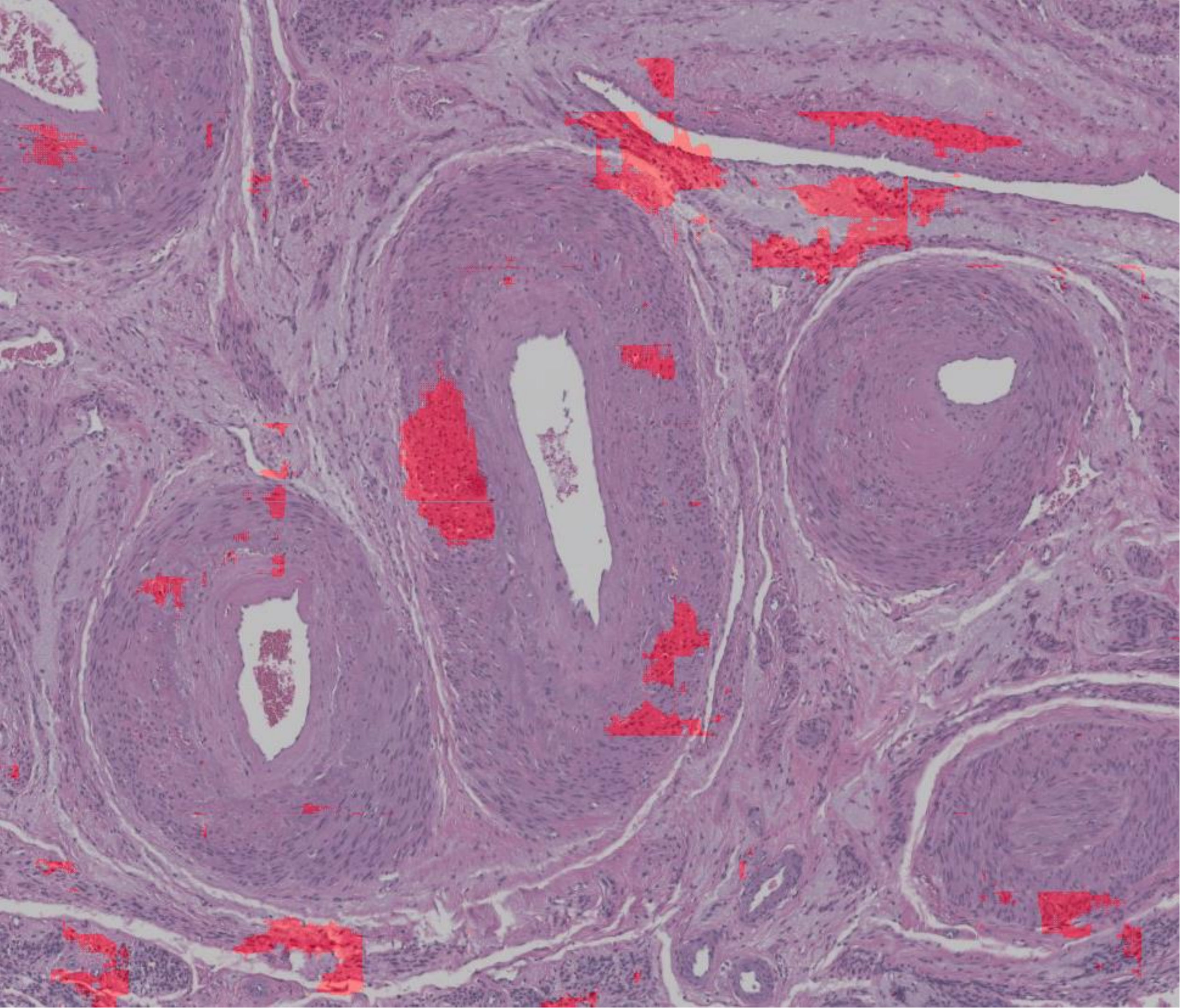,width = 0.47\textwidth}}}
\caption{False positives of our cancer segmentation model. (a,b) Smooth muscle on fallopian tube. (c,d) Colon epithelium. (e,f) Blood vessels. Red regions indicate cancer predicted by the segmentation model.}
\label{fig:FP}
\end{figure*}

To quantitatively analyze ovarian cancer segmentation models, another pathologist who was not involved in training manually generated groundtruth for 14 whole slide images randomly selected from the testing set.
Intersection-over-union (IOU), recall, and precision are used to evaluate segmentation models, where they are defined as:
\begin{equation}
    IOU = \frac{N_{TP}}{N_{TP}+N_{FN}+N_{FP}}
\end{equation}
\begin{equation}
    Recall = \frac{N_{TP}}{N_{TP}+N_{FN}}
\end{equation}
\begin{equation}
    Precision = \frac{N_{TP}}{N_{TP}+N_{FP}}
\end{equation}
where $N_{TP}$, $N_{FN}$, and $N_{FP}$ are the number of true-positive pixels, the number of false-negative pixels, and the number of false-positive pixels, respectively.
Table \ref{tab:seg_results} shows IOU, recall, and precision values for $\mathcal{M}_0$, $\mathcal{M}_1$, $\mathcal{M}_2$, and $\mathcal{M}_3$ where the final model achieved IOU of 0.74, recall of 0.86, and precision of 0.84.
A high precision value and a low recall value from $\mathcal{M}_0$ indicate that the initial model trained by triple-negative breast cancer was not able to segment all high-grade serous ovarian cancer.
After the first iteration by adding papillary patterns for carcinoma, both the IOU value and the recall value were significantly improved.
The second and third iterations had minor updates for correction so the IOU value, the recall value, and the precision value were not significantly improved.
We were able to achieve the highest recall value from the final model indicating the majority of high-grade serous ovarian cancer regions were successfully segmented and heterogeneous morphological patterns would be used to train the \textit{BRCA} classification model.
Figures \ref{fig:img3}, \ref{fig:img2} and \ref{fig:img1} show that our final model can successfully segment ovarian cancers present on three testing whole slide images.
We observed the final model generates false negatives and false positives, shown in Fig. \ref{fig:FN} and \ref{fig:FP}.
Specifically, false negatives we observed were caused by cautery artifact or poor staining.
False positives were caused by smooth muscles on fallopian tube, colon epithelium, or blood vessels which were underrepresented in the training data.
By including normal tissue samples from metastasized cases from other organ types in the training set to further finetune our segmentation model, we expect to reduce false positives.
The segmentation model and code have been released at \href{https://github.com/MSKCC-Computational-Pathology/DMMN-ovary}{https://github.com/MSKCC-Computational-Pathology/DMMN-ovary}.

\begin{table}[ht]
\centering
{
\caption{Intersection-over-union (IOU), recall, and precision for the initial model ($\mathcal{M}_0$), the first model ($\mathcal{M}_1$), the second model ($\mathcal{M}_2$), and the final model ($\mathcal{M}_3$). Note that the initial model is the pretrained breast model \cite{ho2021}. The highest IOU, recall, and precision values are highlighted in bold.}
\begin{tabular}{| c | c | c | c |}
	\hline
	 & IOU & Recall & Precision\\
	\hline
	$\mathcal{M}_0$ & 0.65 & 0.68 & \textbf{0.93}\\
    \hline
    $\mathcal{M}_1$ & \textbf{0.74} & 0.84 & 0.86\\
    \hline
    $\mathcal{M}_2$ & 0.72 & 0.81 & 0.87\\
    \hline
    $\mathcal{M}_3$ & \textbf{0.74} & \textbf{0.86} & 0.84\\
    \hline
\end{tabular}
\label{tab:seg_results}
}
\end{table}

\subsection*{\textit{BRCA} prediction}
Based on cancer segmentation, we trained three \textit{BRCA} classification models in $20\times$, $10\times$, and $5\times$, denoted as $\mathcal{M}_{20\times}$, $\mathcal{M}_{10\times}$, and $\mathcal{M}_{5\times}$, respectively.
During training, the model with the highest area-under-curves (AUCs) on the validation set was selected as the final model.
Table \ref{tab:AUCs} shows AUCs on the validation set and the testing set using the three models in various magnifications where the AUCs were ranging between 0.49 and 0.67 on the validation set and between 0.40 and 0.43 on the testing set.

\begin{table}[ht]
\centering
{
\caption{Area-under-curves (AUCs) of three classification models on the validation set and the testing set.}
\begin{tabular}{| c | c | c | c |}
	\hline
	 & $\mathcal{M}_{20\times}$ & $\mathcal{M}_{10\times}$ & $\mathcal{M}_{5\times}$\\
	\hline
	Validation AUC & 0.49 & 0.65 & 0.67\\
    \hline
    Testing AUC & 0.40 & 0.42 & 0.43\\
    \hline
\end{tabular}
\label{tab:AUCs}
}
\end{table}

\section*{Discussion}
In this paper, we described Deep Interactive Learning (DIaL) which helps annotators to reduce their annotation time to train deep learning-based pixel-wise segmentation models.
Our ovarian cancer segmentation model was able to accurately segment cancer regions presented in H\&E-stained whole slide images with intersection-over-union of 0.74, recall of 0.86, and precision of 0.84.

Cancer segmentation of histologic whole slide images is used to accurately diagnose malignant tissue.
For example, a patch-wise model is designed to identify invasive carcinoma in breast whole slide images \cite{cruz2017}.
Multiple techniques for automated breast cancer metastasis detection in lymph nodes \cite{wang2016,lee2018} has been developed through challenges such as CAMELYON16 \cite{bejnordi2018} and CAMELYON17 \cite{bandi2019}.
For more accurate segmentation, pixel-wise semantic segmentation models such as Fully Convolutional Network (FCN) \cite{long2015}, SegNet \cite{badrinarayanan2017}, and U-Net \cite{ronneberger2015} have been utilized on whole slide images \cite{gecer2018,hermsen2019,seth2019}.
One limitation of these semantic segmentation models is that their input is a patch from a single magnification, where pathologists generally review tissue samples via a microscope in multiple magnifications for cancer diagnosis.
To overcome this challenge in pathology, Deep Multi-Magnification Network (DMMN) utilizing a set of patches from multiple magnifications has been proposed \cite{ho2021}.
In DMMN, patches from $20\times$, $10\times$, and $5\times$ magnifications in a multi-encoder, multi-decoder, multi-concatenation architecture are fully utilized to fuse morphological features from both low magnification and high magnification.
The proposed segmentation network outperformed other single-magnification-based networks.

Cancer segmentation is a critical process not only for diagnosis but also for downstream tasks such as molecular subtyping from H\&E-stained whole slide images.
Molecular status is currently detected by genetic tests but the genetic tests are generally costly and may not be available for all patients.
Deep learning models as screening tools can help patients to get proper treatment from cheap H\&E stains \cite{baxi2021}.
Six mutations were predicted from patches classified as lung adenocarcinoma \cite{coudray2018}.
Microsatellite instability (MSI) status was predicted from patches classified as gastrointestinal cancer \cite{kather2019}.
More molecular pathways and mutations in colorectal cancer were predicted on tumor regions \cite{bilal2021}.
Furthermore, molecular alterations from 14 tumor types were predicted from tumor patches \cite{kather2020}.
\textit{BRCA} mutation \cite{wang2021} and \textit{HER2} status \cite{farahmand2022} in breast cancer were predicted from tumor patches on H\&E-stained images.
To train deep learning models to determine molecular subtypes from H\&E-stained images, manual annotation of tumor regions on whole slide images was required \cite{kather2020,wang2021,farahmand2022}.
To avoid manual annotation of tumor regions for downstream predictions, automated segmentation would be desired.

Alternatively, weakly-supervised learning has been proposed to avoid manual annotation of cancer regions to train cancer segmentation models.
Weakly-supervised learning provides approaches to train classification models by weak labels, such as case-level labels instead of pixel-level labels \cite{campanella2019,lu2021}.
Weakly-supervised learning can be a promising solution for common cancers because it require a large training set representing one whole slide image as one data point.
For example, a weakly-supervised method to predict estrogen receptor status in breast cancer was trained by 2,728 cases \cite{naik2020}.
However, for relatively rare cancers such as high-grade serous ovarian cancer (HGSOC) where the number of cases is limited, weakly-supervised learning can be challenging to detect morphological patterns representing molecular mutations on whole slide images.
In our study, the total number of cases of HGSOC was 609 where only 20\% has \textit{BRCA} status (119 cases).
Therefore, instead of weakly-supervised learning, supervised mutation predictions from cancer segmentation would be more proper especially for rare cancers with limited number of cases.

For automated cancer segmentation, reducing manual annotation time on digitized histopathology images to train segmentation models has been a practical challenge.
Deep learning-based approaches require large quantities of training data with annotations, but pixel-wise annotation for a segmentation model is extremely time-consuming and especially difficult for pathologists with their busy clinical duty.
To reduce annotation burden, approaches to train cell segmentation models with a few scribbles were suggested \cite{lee2020,martinez2021}.
Human-Augmenting Labeling System (HALS) introduced an active learner selecting a subset of training patches to reduce annotation burden for cell annotation \cite{wal2021}.
To segment tissue subtypes, an iterative tool, known as Quick Annotator, was developed to speed up the annotation time within patches extracted from whole slide images \cite{miao2021}.
Patch-level annotation may limit tissue subtypes' field-of-view, potentially causing poor segmentation \cite{bokhorst2019,ho2021}.
Deep Interactive Learning (DIaL) \cite{ho2020} was proposed to efficiently label multiple tissue subtypes in whole slide image-level to reduce time for manual annotation but to have accurate segmentation.
After an initial segmentation training based on initial annotations, mislabeled regions are corrected by annotators and included in the training set to finetune the segmentation model.
As challenging or rare patterns are added during correction, the model can improve its segmentation performance iteratively.
Within 7 hours of manual annotation, the osteosarcoma segmentation model achieved an error rate between its multi-class segmentation predictions and pathologists' manual assessment within an inter-observer variation rate \cite{ho2020}.
In this paper, we further reduced the annotation time by starting from a pretrained segmentation model from a different cancer type to skip the initial annotation step.
In our case, we used a triple-negative breast cancer segmentation model \cite{ho2021} as our initial model and the annotator spent only 3.5 hours to train an accurate ovarian segmentation model.

We desired to predict \textit{BRCA} mutation status from cancer regions from H\&E-stained ovarian whole slide images.
\textit{BRCA} mutation status can determine patients' future treatment but it is currently detected by expensive genetic examinations.
A deep learning-based tool to screen potential \textit{BRCA} cases for genetic examinations from cheap and common H\&E staining would expect to enhance patients' treatment and outcome.
Based on our experiments, we were not able to discover morphological patterns on ovarian cancer indicating \textit{BRCA} mutation.
Several future steps are proposed:
(1) We had a hypothesis that the \textit{BRCA}-related morphological patterns may be shown on cancer regions.
In the future, we may want to expand our hypothesis to non-cancer regions \cite{Brockmoeller22}.
For example, one could train a model from tumor-stroma regions.
(2) We had a hypothesis that the \textit{BRCA}-related morphological patterns may be shown globally.
Therefore, we included all cancer patches to the training set with case-level labels.
If the \textit{BRCA} patterns are shown locally (i.e., some cancer regions may not contain the \textit{BRCA} pattern although its case is labeled as \textit{BRCA}), then weakly-supervised learning approaches \cite{campanella2019,lu2021} may be a better option by increasing the number of cases.
(3) Lastly, cancer morphologies with \textit{BRCA} mutation could be heterogeneous because their mutational spectrum is highly heterogeneous \cite{Hirst18}.
A self-supervised clustering technique could be used to discover multiple morphological patterns of \textit{BRCA} mutation by increasing the number of \textit{BRCA}-mutated cases.

In conclusion, we developed an accurate deep learning-based pixel-wise cancer segmentation model for ovarian cancer.
Especially, by Deep Interactive Learning with a pretrained model from breast cancer, we were able to reduce manual annotation time for training.
Although our study had suboptimal performance on predicting \textit{BRCA} mutation based on morphological patterns, we are confident that our ovarian segmentation model can be used to discover other mutation-related patterns from H\&E-stained images for screening tools to determine treatment and to enhance patient care.

\section*{Data availability}
The segmentation model and code are publicly available at \href{https://github.com/MSKCC-Computational-Pathology/DMMN-ovary}{https://github.com/MSKCC-Computational-Pathology/DMMN-ovary}. 
The data set used and/or analyzed during the current study is available on reasonable request.

\bibliography{references}

\section*{Acknowledgments}
This work was supported by the Warren Alpert Foundation Center for Digital and Computational Pathology at Memorial Sloan Kettering Cancer Center and the NIH/NCI Cancer Center Support Grant P30 CA008748.

\section*{Author contributions}
D.J.H., S.K.P., T.J.F., J.R. conceived the study;
D.J.H. developed the machine learning model and provided statistical analysis in consultation with M.H.C., J.J., M.E.R., and J.R.;
M.H.C. and C.M.V. reviewed and annotated whole slide images;
D.J.H. and J.R. wrote the initial manuscript;
All authors read, edited, and approved the final manuscript.

\section*{Competing interests}
T.J.F. is co-founder, chief scientist and equity holder of Paige.AI.
C.M.V. is a consultant for Paige.AI.
The remaining authors declare no competing interests.

\end{document}